\documentclass[10pt,journal,compsoc]{IEEEtran}

\usepackage{amsfonts}
\usepackage{mathtools}
\usepackage{graphicx}
\usepackage{xspace}
\usepackage{soul}
\usepackage[flushleft]{threeparttable}
\usepackage{multirow}
\usepackage{booktabs}	
\usepackage{subcaption}
\usepackage{enumitem}
\usepackage{caption}
\usepackage{hyperref}
\usepackage{xcolor}

\newcommand{\method}{\mbox{$\mathop{\mathtt{ANT}}\limits$}\xspace}

\newcommand{\methodI}{\mbox{$\mathop{\mathtt{ANT}}\limits$}\xspace}
\newcommand{\methodT}{\mbox{$\mathop{\mathtt{ANT_{e}}}\limits$}\xspace}

\newcommand{\POS}{\mbox{$\mathop{\mathtt{P}}\limits$}\xspace}
\newcommand{\SASRecMP}
{\mbox{$\mathop{\mathtt{SASRec\text{-}P}}\limits$}\xspace}
\newcommand{\SASRec}{\mbox{$\mathop{\mathtt{SASRec}}\limits$}\xspace}
\newcommand{\BERTRec}{\mbox{$\mathop{\mathtt{BERT4Rec}}\limits$}\xspace}
\newcommand{\FDSA}{\mbox{$\mathop{\mathtt{FDSA}}\limits$}\xspace}
\newcommand{\SRec}{\mbox{$\mathop{\mathtt{S3\text{-}Rec}}\limits$}\xspace}

\newcommand{\NPE}{\mbox{$\mathop{\mathtt{NPE}}\limits$}\xspace}
\newcommand{\SASRecP}{\mbox{$\mathop{\mathtt{SASRec\text{++}}}\limits$}\xspace}
\newcommand{\NOVA}{\mbox{$\mathop{\mathtt{NOVA}}\limits$}\xspace}

\newcommand{\UniSRec}{\mbox{$\mathop{\mathtt{UniSRec}}\limits$}\xspace}

\newcommand{\UniSRecI}{\mbox{$\mathop{\mathtt{UniSRec}}\limits$}\xspace}
\newcommand{\UniSRecT}{\mbox{$\mathop{\mathtt{UniSRec_{e}}}\limits$}\xspace}

\newcommand{\RECGUGU}{\mbox{$\mathop{\mathtt{RecGURU}}\limits$}\xspace}
\newcommand{\ZESREC}{\mbox{$\mathop{\mathtt{ZESRec}}\limits$}\xspace}

\newcommand{\NoMoE}{\mbox{$\mathop{\mathtt{ANT \text{-}MoE}}\limits$}\xspace}
\newcommand{\NoText}{\mbox{$\mathop{\mathtt{ANT\text{-} T}}\limits$}\xspace}
\newcommand{\NoPrice}{\mbox{$\mathop{\mathtt{ANT\text{-} P}}\limits$}\xspace}
\newcommand{\NoFusion}{\mbox{$\mathop{\mathtt{ANT\text{-} MF}}\limits$}\xspace}
\newcommand{\NoImage}{\mbox{$\mathop{\mathtt{ANT\text{-} I}}\limits$}\xspace}

\newcommand{\BERT}{\mbox{$\mathop{\mathtt{BERT}}\limits$}\xspace}
\newcommand{\RoBERTa}{\mbox{$\mathop{\mathtt{RoBERTa}}\limits$}\xspace}
\newcommand{\XLM}{\mbox{$\mathop{\mathtt{XLM\text{-}RoBERTa}}\limits$}\xspace}
\newcommand{\SWIN}{\mbox{$\mathop{\mathtt{SWIN}}\limits$}\xspace}
\newcommand{\CLIP}{\mbox{$\mathop{\mathtt{CLIP}}\limits$}\xspace}
\newcommand{\ViT}{\mbox{$\mathop{\mathtt{ViT}}\limits$}\xspace}

\newcommand{\SE}{\mbox{$\mathop{\mathtt{SE}}\limits$}\xspace}

\newcommand{\IRL}{\mbox{$\mathop{\mathtt{IRL}}\limits$}\xspace}

\newcommand{\TOKENIZER}{\mbox{$\mathop{\mathtt{tokenizer}}\limits$}\xspace}
\newcommand{\PROCESSOR}{\mbox{$\mathop{\mathtt{processor}}\limits$}\xspace}

\newcommand{\Scientific}{\mbox{$\mathop{\textsf{\footnotesize{Scientific}}}\limits$}\xspace}
\newcommand{\Pantry}{\mbox{$\mathop{\textsf{\footnotesize{Pantry}}}\limits$}\xspace}
\newcommand{\Instruments}{\mbox{$\mathop{\textsf{\footnotesize{Instruments}}}\limits$}\xspace}
\newcommand{\Arts}{\mbox{$\mathop{\textsf{\footnotesize{Arts}}}\limits$}\xspace}
\newcommand{\Office}{\mbox{$\mathop{\textsf{\footnotesize{Office}}}\limits$}\xspace}

\newcommand{\Food}{\mbox{$\mathop{\textsf{\footnotesize{Food}}}\limits$}\xspace}
\newcommand{\Movies}{\mbox{$\mathop{\textsf{\footnotesize{Movies}}}\limits$}\xspace}
\newcommand{\Cloth}{\mbox{$\mathop{\textsf{\footnotesize{Clothing}}}\limits$}\xspace}
\newcommand{\Home}{\mbox{$\mathop{\textsf{\footnotesize{Home}}}\limits$}\xspace}

\newcommand{\SCRATCH}{\mbox{$\mathop{\mathtt{GB}}\limits$}\xspace}
\newcommand{\FINETUNE}{\mbox{$\mathop{\mathtt{FT}}\limits$}\xspace}
\newcommand{\DSA}{\mbox{$\mathop{\mathtt{ReL}}\limits$}\xspace}

\DeclareMathOperator*{\argmin}{argmin}

\newcommand{\etal}{\textit{et al}.}

\ifCLASSOPTIONcompsoc
  \usepackage[nocompress]{cite}
\else
  \usepackage{cite}
\fi

\begin{document}

\title{Multi-modality Meets Re-learning: \\
Mitigating Negative Transfer in Sequential Recommendation}


\author{Bo~Peng,
    Srinivasan~Parthasarathy,~\IEEEmembership{Fellow,~IEEE,}
        and~Xia~Ning$^*$,~\IEEEmembership{Member,~IEEE}
\IEEEcompsocitemizethanks{
\IEEEcompsocthanksitem Bo Peng
is with the Department
of Computer Science and Engineering, The Ohio State University, Columbus,
OH, 43210.\protect\\
E-mail: peng.707@buckeyemail.osu.edu
\IEEEcompsocthanksitem Srinivasan Parthasarathy 
is with the Department of Computer Science and Engineering, 
and the Translational Data Analytics Institute,
The Ohio State University, Columbus, OH, 43210.\protect\\
E-mail:  srini@cse.ohio-state.edu
\IEEEcompsocthanksitem Xia Ning 
is with the Department of Biomedical Informatics, 
the Department of Computer Science and Engineering, 
and the Translational Data Analytics Institute,
The Ohio State University, Columbus, OH, 43210.\protect\\
E-mail: ning.104@osu.edu
\IEEEcompsocthanksitem $^*$Corresponding author
}
\thanks{paper received April 19, 2005; revised August 26, 2015.}}

\markboth{Journal of \LaTeX\ Class Files,~Vol.~14, No.~8, August~2015}%
{Shell \MakeLowercase{\textit{et al.}}: Bare Demo of IEEEtran.cls for Computer Society Journals}
%

%


\IEEEtitleabstractindextext{%
\begin{abstract}

Learning effective recommendation models from sparse user interactions represents a fundamental challenge in developing sequential recommendation methods.
Recently, pre-training-based methods have been developed to tackle this challenge.
%
%
Though promising, in this paper, we show that existing methods suffer from the notorious negative transfer issue, 
where the model adapted from the pre-trained model results in worse performance compared to the model learned from scratch in the task of interest (i.e., target task).
To address this issue, we develop a method, denoted as \method, for transferable sequential recommendation.
%
\method mitigates negative transfer by 
1) incorporating multi-modality item information, including item texts, images and prices,
 to effectively learn more transferable knowledge from related tasks (i.e., auxiliary tasks); 
and 
2) better capturing task-specific knowledge in the target task using a re-learning-based adaptation strategy.
We evaluate \method against eight state-of-the-art baseline methods on five target tasks.
Our experimental results demonstrate that 
\method does not suffer from the negative transfer issue on any of the target tasks.
The results also demonstrate that \method substantially outperforms 
baseline methods in the target tasks 
with an improvement of as much as 15.2\%.
%
%
%
Our analysis highlights the superior effectiveness of our re-learning-based strategy compared to fine-tuning on the target tasks.
\end{abstract}

}
\maketitle

\IEEEdisplaynontitleabstractindextext
\IEEEpeerreviewmaketitle

\section{Introduction}
\label{sec:intro}

\IEEEPARstart{S}{equential}
recommendation aims to 
leverage users' historical interactions 
to identify and recommend the next item of their interest.
Due to the explosion in the number of available items, 
it has become indispensable in various applications  
such as online retail~\cite{he2016ups} 
and video streaming~\cite{belletti2019quantifying}, 
drawing increasing attention from the research community.
A fundamental challenge in developing modern sequential recommendation methods is to learn effective recommendation models 
from the sparse user interaction data in the task of interest, referred to as the target task.
Recently, pre-training-based recommendation methods~\cite{hou2022towards,ding2021zero}
have been developed to tackle this challenge.
These methods 
learn transferable knowledge from the rich interaction
data in other related tasks, referred to as auxiliary tasks, 
via pre-training, and transfer the learned knowledge to improve recommendations for the target task.

To enable an effective transfer learning in the target task,  
two significant challenges need to be addressed: 
1) learning transferable knowledge from auxiliary tasks applicable and beneficial for recommendations in the target task; 
and 2) capturing task-specific knowledge in the target task.
Current methods~\cite{hou2022towards,ding2021zero} primarily learn transferable knowledge using item texts.
However, as will be demonstrated in our analysis (Section~\ref{sec:exp:domain}), 
item texts in auxiliary tasks and the target task could 
exhibit remarkable differences in terms of 
wording, style and vocabulary.
Consequently, limited transferable knowledge could be 
learned from item texts.
In addition, 
existing methods~\cite{hou2022towards} learn the task-specific knowledge in the target task via fine-tuning.
However, starting from parameters pre-trained on auxiliary tasks, the fine-tuned models may struggle to capture the task-specific knowledge in the target task since the target task could be considerably different from the auxiliary ones.

Due to the above limitations, 
as will be shown in our analysis (Section~\ref{sec:exp:transferibility}),  
current methods could suffer from the notorious negative transfer issue so that 
the model adapted from the pre-trained model (i.e., adapted model) underperforms the model learned from scratch (i.e., ground-built model) in the target task.
In this paper, we develop a method, 
denoted as \method, to address the negative transfer issue in transferable sequential recommendation.
%
In order to effectively learn more transferable knowledge from auxiliary tasks, \method not only utilizes item texts but also incorporates item images and prices.
As will be shown in our analysis (Section~\ref{sec:exp:domain}), item images and prices exhibit better transferability over item texts across tasks.
\method also employs a re-learning-based strategy to 
better capture the task-specific knowledge in the target task. 
Specifically, instead of fine-tuning, 
\method relearns certain parameters in 
the pre-trained model during adaptation
to enable a more task-specific model in the target task.

We conduct a comprehensive evaluation to compare \method 
against eight state-of-the-art baseline methods in five target tasks.
Our experimental results demonstrate that \method does not suffer from the negative transfer issue on any of the five target tasks.
The results also demonstrate that in the five target tasks, 
\method significantly outperforms baseline methods, 
with an improvement of up to 15.2\%.
Our ablation study demonstrates that item texts, images and prices all significantly contribute to the recommendation performance of \method.
%
%
%
%
%
For better reproducibility, 
we release the processed data, source code and pre-trained recommendation model in \url{https://github.com/ninglab/ANT}.
%

\section{Related Work}
\label{sec:related}

In the last few years, numerous sequential  recommendation methods have been developed, which leverage neural networks (e.g., recurrent neural networks) and attention mechanisms.
For example, Hidasi~\etal~\cite{hidasi2015session} developed the first gated recurrent \mbox{units-based} 
methods to estimate users' preferences in a recurrent manner.
Nguyan~\etal~\cite{nguyen2018npe} developed a neural personalized embedding model (\NPE) 
in which multilayer perceptrons are utilized to capture the non-linear relations among items.
Recently, self-attention mechanisms have also been largely 
employed in developing sequential recommendation methods.
Kang~\etal~\cite{kang2018self} developed the first self-attention-based sequential recommendation method \SASRec, 
which stacks multiple self-attention layers to estimate users' preferences in an iterative manner.
Inspired by \SASRec, Sun~\etal~\cite{sun2019bert4rec} further utilized the cloze objective~\cite{devlin2018bert} to enable 
a self-attention-based bidirectional sequence encoder to better capture users' preferences.
Zhang~\etal~\cite{zhang2019feature} developed \FDSA in which two self-attention networks are employed to capture the dynamics among items and item texts.
Zhou~\etal~\cite{zhou2020s3} developed a self-supervised attentive method \SRec in which the mutual information 
between item embeddings and item attribute embeddings are maximized to enable more expressive representations.
Liu~\etal~\cite{liu2021noninvasive} developed \NOVA, a non-invasive self-attentive model to fuse multi-modality item information for better recommendation.
Rashed~\etal~\cite{rashed2022carca} developed \SASRecP which integrates multi-modality item information (e.g., texts and images) to improve the performance of \SASRec.

To deal with the limited user interaction data in single recommendation tasks, recently, 
transfer-learning-based sequential recommendation methods are developed to transfer knowledge from auxiliary tasks to the target task for better recommendations.
Li~\etal~\cite{li2022recguru} developed \RECGUGU, which utilizes a self-attentive autoencoder to generate 
transferable user representations recommendation.
Ding~\etal~\cite{ding2021zero} developed \ZESREC, which utilizes 
item text embeddings from pre-trained language models as transferable representations of items, 
and employed the pre-training and fine-tuning framework to transfer knowledge across tasks.
Similarly to \ZESREC, Hou~\etal~\cite{hou2022towards} also 
leveraged item text embeddings as transferable item representations and developed \UniSRec.
However, different from that in \ZESREC, in \UniSRec, 
an embedding transformation module is included to transform the text embeddings and optimize the recommendation performance in the target task.
%
%

%
%

\section{Definitions and Notations}
\label{sec:definition}

%

%
In this paper, we denote a single item as $v$. 
For the $i$-th item $v_i$, we represent its text, image and price as $v_i^t$, $v_i^m$ 
and $v_i^p$, respectively.
We denote a user as $u$.
For the $j$-th user $u_j$ in a task, 
we denote her/his historical interactions as a sequence 
$S_j = \{v_{s_1}(j), v_{s_2}(j), \cdots\}$, 
where $v_{s_t}(j)$ is the $t$-th interacted item in $S_j$.
Given $S_j$, the next item that $u_j$ will interact with is referred to as the ground-truth next item, denoted as $v_{g_j}$.
The goal of \method is to correctly recommend $v_{g_j}$ for $u_j$.
When no ambiguity arises, we will eliminate $j$ in $S_j$, $v_{s_t}(j)$ and $v_{g_j}$.
We list the key definitions used in this paper as follows:
\begin{itemize}[leftmargin=*]
    \item Recommendation task: We consider recommending items 
    within a single category as a recommendation task (e.g., movie recommendation, clothes recommendation), following the literature~\cite{li2022recguru,ding2021zero,hou2022towards}. 
    This is because users could exhibit different behavior patterns in interacting items of different categories.
    \item Target task: The recommendation task of interest is referred to as the target task, denoted as $T^*$. 
    We denote the set of all the items in $T^*$ as $\mathcal{V}^*$.
    \item Auxiliary task: The recommendation task used to learn transferable knowledge is referred to as an auxiliary task, denoted as $T$.
    We denote the set of all the auxiliary tasks as $\mathcal{A} = \{T_1, T_2, \cdots\}$, and the set of all the items in the $k$-th auxiliary task $T_k$ as $\mathcal{V}_{k}$.
    \item User intent modeling: we consider the process of estimating users' preferences via their interacted items as user intent modeling following \cite{chang2022latent}.
\end{itemize}
We denote matrices, scalars 
and row vectors using uppercase, lowercase,
and bold lowercase letters, respectively.
%

\section{Method}
\label{sec:method}

\begin{figure*}[h]
\centering
\includegraphics[width=0.8\linewidth]{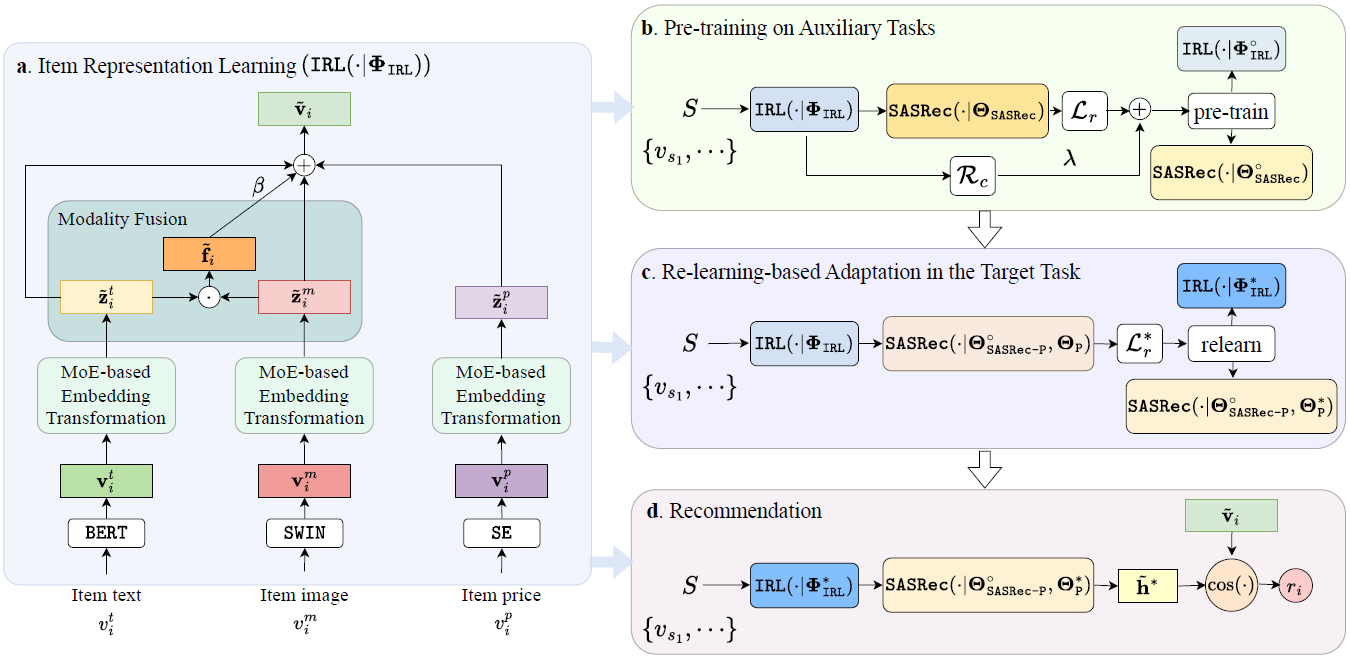}
\caption{
\method Model Architecture.
Fig. 1a: \method utilizes item texts, images and prices together to generate item embeddings;
Fig. 1b: \method pre-trains the parameters in \IRL (i.e., $\boldsymbol{\Phi}_{\scriptsize{\IRL}}^\circ$) and the user intent model (i.e., $\boldsymbol{\Theta}_{\scriptsize{\SASRec}}^\circ$) to learn transferable knowledge from auxiliary tasks;
Fig. 1c: \method relearns the parameters in \IRL (i.e., $\boldsymbol{\Phi}_{\scriptsize{\IRL}}^*$) and the position embeddings ($\boldsymbol{\Theta}_{\scriptsize{\POS}}^*$) to capture the task-specific knowledge in the target task;
Fig. 1d: \method generates recommendations using pre-trained and relearned parameters.
}
\vspace{-10pt}
\label{fig:architecture}
\end{figure*}

Figure~\ref{fig:architecture} presents the \method architecture, and the pre-training, adaptation and recommendation processes in \method.
\method transfers knowledge from auxiliary tasks 
to the target task via pre-training and adaptation.
In particular, we pre-train \method on auxiliary tasks 
to capture transferable knowledge for recommendations.
We assume such knowledge is applicable to the target task, and thus, adapt the pre-trained model to the target task to improve the recommendation performance.
To address the negative transfer issue,
\method utilizes item texts, images and prices 
together to learn more transferable knowledge
from auxiliary tasks.
In addition, 
\method employs a re-learning-based adaptation strategy to enable a more task-specific model in the target task.
In what follows, we first present the item representation learning and user intent modeling in \method in Section~\ref{sec:method:representation} and Section~\ref{sec:method:intent}, respectively.
Then, we present the pre-training process in \method in Section~\ref{sec:method:pretrain} 
and the adaptation strategy in \method in Section~\ref{sec:method:finetuning}.

\subsection{Item Representation Learning (\IRL)}
\label{sec:method:representation}

Existing methods~\cite{hou2022towards,ding2021zero} learn item representations using only item texts.
However, as will be shown in our own analysis (Section~\ref{sec:exp:domain}), 
the styles, vocabularies and wording of item texts in auxiliary tasks and the target task could be significantly different.
Consequently, based exclusively on item texts, we may learn limited transferable knowledge from auxiliary tasks.
In order to capture more transferable knowledge, 
in \method, 
besides item texts, we also incorporate item images and prices when generating item representations.
Specifically, in \method, we utilize a fixed pre-trained language model (e.g., \BERT), a fixed pre-trained vision model (e.g., \SWIN) and the sinusoidal encoding (\SE), all of which are recommendation-independent,  
to generate embeddings from item texts, images and prices, respectively.
Then, we transform the generated embeddings to better adapt their semantics for user intent modeling.
We fix the pre-trained language model and vision model for better efficiency following the literature~\cite{hou2022towards,ding2021zero}.


\subsubsection{Text Embedding Generation}
\label{sec:method:representation:text}

In \method, we use \BERT~\cite{devlin2018bert} to 
generate text embeddings. 
%
We also tried 
other pre-trained language models such as 
\RoBERTa~\cite{liu2019roberta} 
and \XLM~\cite{XLM}
but empirically observed that \BERT has the best performance.
Given the item texts, the pre-trained \BERT model 
downloaded from Hugging Face~\footnote{\url{https://huggingface.co/models}}
and the associated tokenizer, 
we generate the text embeddings as follows:
\begin{equation}
    \label{eqn:text}
    \mathbf{v}_i^t = \BERT \left( \TOKENIZER \left( v_i^t \right) \right),
\end{equation}
where $v_i^t$ and $\mathbf{v}_i^t$ are the text and text embedding of item $v_i$, respectively. 

\subsubsection{Image Embedding Generation}
\label{sec:method:representation:image}

In \method, we utilize \SWIN~\cite{liu2021swin} to generate image embeddings
since we empirically observed that it outperforms other popular pre-trained vision models such as \CLIP~\cite{radford2021learning} 
and \ViT~\cite{dosovitskiy2020image}. 
Particularly, given the item image, the pre-trained \SWIN model from Hugging Face
and the associated image processor, we generate image embeddings as follows:
\begin{equation}
    \label{eqn:image}
    \mathbf{v}_i^m = \SWIN \left( \PROCESSOR \left( v_i^m \right) \right),
\end{equation}
where $v_i^m$ and $\mathbf{v}_i^m$ are the image and image embedding of item $v_i$, respectively.

\subsubsection{Price Embedding Generation}
\label{sec:method:representation:price}

Embeddings for numerical variables such as prices
should be monotonic, transnationally invariant and symmetric in terms of their proximity.
As proved in Wang~\cite{wang2020position}, 
by using an appropriate frequency 
coefficient, 
\SE~\cite{vaswani2017attention} 
could generate embeddings that
guarantee the above properties.
Therefore, in \method, 
we utilize \SE to generate price embeddings as follows:

\begin{equation}
\label{eqn:price}
    \mathbf{v}_{i}^{p} = 
    \left[ \cdots, \sin \left( \omega_j v_{i}^{p} \right), 
    \cos \left( \omega_j v_{i}^{p} \right),
    \cdots \right], \quad
    \omega_j = {\omega} ^{-\frac{2j}{d_p}}, 
\end{equation}
where $v_{i}^{p}$ and $\mathbf{v}_{i}^{p}$ is 
the price and price embedding of item $v_i$, respectively;
$\omega_j$ is the coefficient in the $2j$ and $2j$+1-th dimension of $\mathbf{v}_{i}^{p}$;
$\omega$ is the frequency coefficient in \SE, 
and $d_p$ is the dimension of price embeddings.
%

\subsubsection{Embedding Transformation}
\label{sec:method:representation:semantic}

%
We transform embeddings generated from recommendation-independent \BERT, \SWIN and \SE to capture more recommendation-specific information.
Particularly, we 
use the mixture-of-expert (MoE) mechanism
for the embedding transformation.
During the transformation, we use a projection layer to project the embeddings to spaces that could consist of recommendation-specific information.
We consider the projected embeddings as experts and leverage a routing layer to adaptively aggregate the experts
to generate transformed embeddings.
In \method, 
we use parametric whitening~\cite{huang2021whiteningbert,hou2022towards,su2021whitening} for texts and images, 
and the traditional linear projection for prices. 

\paragraph*{Parametric Whitening for Texts and Images} 

Previous study~\cite{hou2022towards} utilizes parametric whitening to project text embeddings and demonstrate superior performance in recommendation tasks.
In \method, we generalize this idea to images and perform the projection 
of text and image embeddings as follows:
\begin{equation}
    \label{eqn:parametric}
    \tilde{\mathbf{v}}^x_i(k) = \left( \mathbf{v}_i^x - \mathbf{b}^x_k \right ) W^x_k,
\end{equation}
where $x$ denotes the modality, that is $x = t$ or $m$; $\mathbf{v}_i^x$
is item $v_i$'s embedding for modality $x$;
$\mathbf{b}^x_k$ is a learnable 
shift parameter for texts or images, 
and $W_k^x$ 
is a learnable projection parameter 
in the $k$-th projection head, $k \in \{1, 2, \cdots, n_h\}$; 
$\tilde{\mathbf{v}}^x_i(k)$
is the projected embedding 
from the $k$-th head.

\paragraph*{Linear Projection for Prices}
We utilize the widely used linear 
projection for the transformation of price embeddings as follows:
\begin{equation}
    \label{eqn:parametric}
    \tilde{\mathbf{v}}_i^p(k) = \mathbf{v}_i^p W_k^p + \mathbf{b}_k^p,
\end{equation}
where $\mathbf{v}_i^p$ is the price embedding of item $v_i$, 
and $W_k^p$ and $\mathbf{b}_k^p$ are learnable parameters in the $k$-th projection head. We empirically observed that linear projection outperforms parametric whitening for price embeddings.

\paragraph*{Aggregation via Gaussian Routing}
  
We learn weights following Gaussian distributions 
to aggregate the projected embedding 
$\tilde{\mathbf{v}}_i^x(k)$ ($x=t$, $m$ or $p$) for the $i$-th item as follows:
\begin{equation}
    \label{eqn:aggregate}
    \tilde{\mathbf{z}}_i^x = \text{Softmax} (\boldsymbol{\alpha}_i^x) [\tilde{\mathbf{v}}_i^x(1); \tilde{\mathbf{v}}_i^x(2); \cdots; \tilde{\mathbf{v}}_i^x(n_h)],
\end{equation}
where Softmax is employed to normalize the learned weights 
\\
\mbox{$\boldsymbol{\alpha}_i^x \sim \mathcal{N}(\boldsymbol{\mu}_i^x, \text{diag}((\boldsymbol{\sigma}_i^x)^2))$}, 
and $\tilde{\mathbf{z}}_i^x$ ($x = t$, $m$ or $p$) 
is the transformed embedding of item $v_i$ from modality $x$.
We parameterize the mean $\boldsymbol{\mu}_i^x$ and standard deviation $\boldsymbol{\sigma}_i^x$ of the Gaussian distribution as follows:
\begin{equation}
    \label{eqn:mean}
    \boldsymbol{\mu}_i^x = \mathbf{v}_i^x B^x 
    \; \text{and} \;
    \boldsymbol{\sigma}_i^x = \text{Softplus} (\mathbf{v}_i^x U^x), 
\end{equation}
where $B^x$ ($x=t$, $m$ or $p$) and 
$U^x$ are learnable parameters;
Softplus is the activation function to 
ensure a positive standard deviation.
Compared to the widely used linear 
and cosine~\cite{chi2022representation} routing, 
the Gaussian routing captures the uncertainty in the weight generation process by modeling it as the variance of the Gaussian distribution, 
and thus, could improve the model generalizability~\cite{li2022uncertainty}.

\subsubsection{Modality Fusion}
\label{sec:method:fusion}

The synergy among item modalities (e.g., texts and images) have been demonstrated~\cite{radford2021learning} useful for recommendation.
For example, with an image of a can and the word ``cat" in the text together, 
we could identify the
item as cat can food 
rather than mistaking it as fish or soda cans
based on the item image alone.
In \method, 
%
given the transformed text and image embedding 
$\tilde{\mathbf{z}}_i^t$ and $\tilde{\mathbf{z}}_i^m$ of item $v_i$,
we capture the synergy between the text and image as follows: 
\begin{equation}
    \label{eqn:fusion}
    \tilde{\mathbf{f}}_i = \tilde{\mathbf{z}}_i^t \odot \tilde{\mathbf{z}}_i^m, 
\end{equation}
where $\tilde{\mathbf{f}}_i$ is an embedding representing  
the synergy between the text and image of item $v_i$;
$\odot$ is the Hadamard product (i.e., element-wise product).
Compared to the widely used Transformer-based approach~\cite{xie2022decoupled}, 
our Hadamard product-based method is significantly more efficient and could still effectively capture the synergy among the modalities.
Note that, item texts and images both reflect intrinsic properties of items (e.g., category, functionality), 
while item prices could be substantially affected by external factors (e.g., seasonal, festival).
Therefore, we only consider synergies between item texts and images, not including prices.

In \method, we leverage item texts, item images, their synergy and item prices to generate comprehensive item embeddings as follows:
\begin{equation}
    \label{eqn:item}
    \tilde{\mathbf{v}}_i = \IRL(v_i^t, v_i^m, v_i^p | \boldsymbol{\Phi}_{\scriptsize{\IRL}}, \beta) =
    \tilde{\mathbf{z}}_i^t + \tilde{\mathbf{z}}_i^m + \tilde{\mathbf{z}}_i^p + \beta \tilde{\mathbf{f}}_i,
\end{equation}
where $\tilde{\mathbf{v}}_i$ is the comprehensive item embedding of $v_i$. 
We introduce a hyper-parameter $\beta \in (0, 1)$ to 
control the weight of $\tilde{\mathbf{f}}_i$ 
and mitigate the inflation of 
common information from text and image into $\tilde{\mathbf{v}}_i$.
In Equation~\ref{eqn:item}, 
$\boldsymbol{\Phi}_{\scriptsize{\IRL}}$ is the set of all the learnable parameters for item representation learning:
\begin{equation}
    \label{eqn:phi}
    \boldsymbol{\Phi}_{\scriptsize{\IRL}} = \{\mathbf{b}_k^x, W_k^x, B^x, U^x | x \in \{t, m, p\}, k \in \{1, \cdots, n_h\}\}.
\end{equation}
%

\subsection{User Intent Modeling}
\label{sec:method:intent}

Following existing pre-training-based methods~\cite{wang2022transrec,hou2022towards}, in this paper, we utilize the widely used \SASRec model for user intent modeling in \method.
However, it should be noted that \method serves as a general framework that is also compatible with other user intent modeling methods.
\SASRec is a self-attention-based~\cite{vaswani2017attention} 
method, which learns temporal patterns using position embeddings, and 
stacks multiple self-attention layers 
to aggregate interacted items and 
estimate users' preference in an iterative manner.
Specifically, given the interaction 
sequence $S = \{s_1, s_2, \cdots\}$, 
we stack the embedding of interacted items  
to generate an embedding matrix for $S$ as follows:
\begin{equation}
    \label{eqn:matrix}
    M = [\tilde{\mathbf{v}}_{s_1}; \tilde{\mathbf{v}}_{s_2}; \cdots], 
\end{equation}
where $\tilde{\mathbf{v}}_{s_i}$ is the embedding of 
item $s_i$ (Equation~\ref{eqn:item}).
%
%
%
With $M$, we generate the 
hidden representation of users' preference $\tilde{\mathbf{h}}$ via \SASRec:
\begin{equation}
    \label{eqn:intent}
    \tilde{\mathbf{h}} = \SASRec(M | \boldsymbol{\Theta}_{\scriptsize{\SASRecMP}}, 
    \boldsymbol{\Theta}_{\scriptsize{\POS}}),
\end{equation}
where $\boldsymbol{\Theta}_{\scriptsize{\SASRecMP}}$ represents all the learnable parameters in \SASRec excluding the position embeddings, 
and $\boldsymbol{\Theta}_{\scriptsize{\POS}}$ represents all the learnable parameters in position embeddings.
We use \SASRec for the user intent modeling in both pre-training on auxiliary tasks (Section~\ref{sec:method:pretrain}) and adaptation in the target task (Section~\ref{sec:method:finetuning}).

\subsection{Pre-training on Auxiliary Tasks}
\label{sec:method:pretrain}

The user interaction data in the target task 
(i.e., movie recommendation)
is generally sparse.
The sparse data may not carry enough information to enable effective recommendation models. 
To leverage richer data from many auxiliary tasks, 
in \method, we learn and transfer knowledge from auxiliary tasks 
to the target task to improve the recommendation performance in the target task.
Particularly, motivated by the success of 
model pre-training in transfer learning~\cite{devlin2018bert,liu2021swin,radford2021learning},
we pre-train a recommendation model 
on auxiliary tasks. 
%
%
%
%
In this Section, we present the loss function and regularization used in our pre-training in detail.

\subsubsection{Item Recommendation Loss}
\label{sec:method:pretrain:recommendation}

We introduce an item recommendation loss in pre-training 
to learn a pre-trained model 
from all the auxiliary tasks.
%
%
We define the negative 
log likelihood of correctly recommending the next item 
as the recommendation loss as follows:
\begin{equation}
    \label{eqn:rec}
    \begin{aligned}
        \mathcal{L}_r &= - \sum_{T_k \in \mathcal{A}}\,
        \space \sum_{(S_j, v_{g_j}) \in T_k} \log(p_j), \\
        \text{where}  \quad  p_j &= \frac{\exp(\cos(\tilde{\mathbf{h}}_j, \tilde{\mathbf{v}}_{g_j}) / \tau)}
        {\sum_{v_{i^\prime} \in \cup \mathcal{V}_k} 
        \exp(\cos(\tilde{\mathbf{h}}_j, \tilde{\mathbf{v}}_{i^\prime}) / \tau)},
    \end{aligned}
\end{equation}
where $v_{g_j}$ is the ground-truth next item of the $j$-th user's interest in an auxiliary task;
$p_j$ is the probability of correctly recommending $v_{g_j}$;
$\tilde{\mathbf{h}}_j$ is the preference of the $j$-th user in an auxiliary task following \SASRec (Equation~\ref{eqn:intent}); 
$\tilde{\mathbf{v}}_{g_j}$ is the embedding of $v_{g_j}$ (Equation~\ref{eqn:item});
$\mathcal{V}_k$ is the set of all the items in the $k$-th auxiliary task;
$\cup \mathcal{V}_k$ is the union of items in all the auxiliary tasks;
$\cos(\cdot)$ is the cosine similarity.
%

\subsubsection{Text-image Alignment Regularization}
\label{sec:method:pretrain:contrastive}

Inspired by \CLIP~\cite{radford2021learning}, 
we introduce a text-image alignment regularization to 
capture the potential complementarity between texts and images.
Specifically, given the set of all the items 
in auxiliary tasks (i.e., $\cup \mathcal{V}_k$), 
we formulate the text-image alignment regularization 
using a negative log likelihood as:
\begin{equation}
    \vspace{-5pt}
    \label{eqn:contrastive}
    \begin{aligned}
        \mathcal{R}_c &= - \sum_{v_i \in \cup \mathcal{V}_k} \log(q_i), \\
        \text{where}  \quad q_i &= \frac{\exp(\cos(\tilde{\mathbf{z}}_{i}^m, \tilde{\mathbf{z}}_{i}^t) / \tau)}
        {\sum_{v_{i^\prime} \in \cup \mathcal{V}_k} \exp(\cos(\tilde{\mathbf{z}}_{{i}}^m, \tilde{\mathbf{z}}_{{i^\prime}}^t) / \tau)},
    \end{aligned}
\end{equation}
where $q_i$ is the probability of 
$v_i$'s text and image carrying the same information;
$\tilde{\mathbf{z}}_{{i}}^t$ and $\tilde{\mathbf{z}}_{{i}}^m$ is the transformed embedding of the text and image of item $v_i$, respectively (Equation~\ref{eqn:aggregate});
$\tau \in [0, 1]$ is a hyper-parameter to scale the similarities.
%

\subsubsection{Pre-training}
\label{sec:method:pretrain:training}

Using all the interactions in all the auxiliary tasks, 
we pre-train the model from auxiliary tasks and optimize the parameters in \method as follows:
\begin{equation}
    \begin{aligned}
    \label{eqn:multi}
    (\boldsymbol{\Phi}_{\scriptsize{\IRL}}^\circ, \boldsymbol{\Theta}_{\scriptsize{\SASRec}}^\circ) 
    &= \argmin \limits_{\boldsymbol{\Phi}_{\scriptsize{\IRL}}, \boldsymbol{\Theta}_{\scriptsize{\SASRec}}} 
    \mathcal{L}_r\!+\!\lambda \mathcal{R}_c, \\
    \text{where } \mathcal{L}_r\!+\!\lambda \mathcal{R}_c 
    &= \!\!-\!\!\sum_{T_k \in \mathcal{A}} \;\;\;\;\;
    \sum_{\mathclap{\substack{(S_j, g_j) \in T_k}}} \log(p_j)\!-\! 
    \lambda \sum_{\mathclap{\substack{v_i \in \cup \mathcal{V}_k}}}
    \log(q_i),
    \end{aligned}
\end{equation}
where $\boldsymbol{\Phi}_{\scriptsize{\IRL}}^\circ$ (Equation~\ref{eqn:phi}) is the set of pre-trained parameters for item representation learning that generates $\tilde{\mathbf{v}}_i^x$ (Equation~\ref{eqn:item});
$\boldsymbol{\Theta}_{\scriptsize{\SASRec}}^\circ = \{\boldsymbol{\Theta}_{\scriptsize{\SASRecMP}}^\circ, \boldsymbol{\Theta}_{\scriptsize{\POS}}^\circ \}$ 
is the set of pre-trained parameters in user intent modeling 
that generates $\tilde{\mathbf{h}}$ (Equation~\ref{eqn:intent});
$\lambda$ is the coefficient for the regularization.
We set $\lambda$ as $1e$-$3$ in our pre-training for simplicity.

\subsection{Re-learning-based Adaptation in the Target Task}
\label{sec:method:finetuning}

Existing work~\cite{hou2022towards,ding2021zero} fine-tunes the pre-trained model within the target task to capture target-task-specific knowledge.
However, the pre-trained model captures general knowledge from all the auxiliary tasks.
Directly fine-tuning the pre-trained model may 
end up with a local optimum that is not 
specific enough to capture the unique user behavior patterns in the target task.
To address this issue,
we develop a re-learning-based adaptation strategy in which instead of fine-tuning, 
we relearn the parameters for item representation learning in $\boldsymbol{\Phi}_{\scriptsize{\IRL}}$ (Equation~\ref{eqn:phi}) from scratch in the target task.
Intuitively, different tasks require prioritizing different item properties for effective recommendation (e.g., styles in clothes recommendation, casting in movie recommendation).
To this end, 
highly task-specific parameters are desired for item representation learning.
In addition, we also relearn the position embeddings in $\boldsymbol{\Theta}_{\scriptsize{\POS}}$ (Equation~\ref{eqn:intent}) to capture the unique temporal patterns in the target task.
Compared to fine-tuned parameters, the relearned ones are more specific to the target task, and thus, could better capture the task-specific knowledge desired.
During adaptation,  
we also fix pre-trained parameters in user intent modeling $\boldsymbol{\Theta}_{\scriptsize{\SASRecMP}}^\circ$ since we empirically found that updating these parameters does not benefit the recommendation performance in the target task.
Note that, although we relearn $\boldsymbol{\Phi}_{\scriptsize{\IRL}}$ in the target task, 
we still implicitly use $\boldsymbol{\Phi}_{\scriptsize{\IRL}}^\circ$ during adaptation 
because they are used in $\boldsymbol{\Theta}_{\scriptsize{\SASRecMP}}^\circ$.
%
%
In summary, during adaptation, \method relearns the parameters in $\boldsymbol{\Phi}_{\scriptsize{\IRL}}$ and $\boldsymbol{\Theta}_{\scriptsize{\POS}}$ to capture the target-task-specific knowledge, 
while \method fixes the pre-trained parameters in $\boldsymbol{\Theta}_{\scriptsize{\SASRecMP}}$ 
to leverage the transferable knowledge learned from auxiliary tasks. 

\paragraph*{\method Training in the Target Task}
We relearn \method in the target task by minimizing the negative log likelihood of correctly recommending the next item in the target task as follows:
\begin{equation}
    \label{eqn:loss:target}
    \begin{aligned}
        \min \limits_{\boldsymbol{\Phi}_{\scriptsize{\IRL}}^*, \boldsymbol{\Theta}_{\scriptsize{\POS}}^*} 
        \mathcal{L}_r^*              &= - \sum_{\mathclap{(S_j, v_{g_j}) \in T^*}} \log(p_j^*), \\
        \text{where} \quad p_j^* &= \frac{\exp(\cos(\tilde{\mathbf{h}}_j^*, \tilde{\mathbf{v}}_{g_j}^*) / \tau)}
        {\sum_{v_{i^\prime} \in \mathcal{V}^*} 
        \exp(\cos(\tilde{\mathbf{h}}_j^*, \tilde{\mathbf{v}}_{i^\prime}^*) / \tau)}, \\
        \tilde{\mathbf{v}}_i^*      & = \IRL(v_i^t, v_i^m, v_i^p | \boldsymbol{\Phi}_{\scriptsize{\IRL}}^*), \\
        \tilde{\mathbf{h}}^*         & = \SASRec(M^*|\boldsymbol{\Theta}_{\scriptsize{\SASRecMP}}^\circ, \boldsymbol{\Theta}_{\scriptsize{\POS}}^*),
    \end{aligned}
\end{equation}
where $T^*$ is the target task; 
$\mathcal{V}^*$ is the set of all the items in the target task;
$\tilde{\mathbf{v}}_{i^\prime}^*$ is the embedding of 
$v_{i^\prime}$ in the target task;
$\tilde{\mathbf{h}}_j^*$ is the preference 
of the $j$-th user in the target task;
$M^*$ is a matrix containing embeddings of 
the interacted items in $S_j$ (Eqauation~\ref{eqn:matrix}).
%
%
%
As shown in Equation~\ref{eqn:loss:target}, during adaptation, 
we relearn the parameters for item representation learning (i.e., $\boldsymbol{\Phi}_{\scriptsize{\IRL}}^*$), and temporal patterns modeling 
(i.e., $\boldsymbol{\Theta}_{\scriptsize{\POS}}^*$)
while fixing pre-trained parameters in $\boldsymbol{\Theta}_{\scriptsize{\SASRecMP}}^\circ$ (Equation~\ref{eqn:intent}).
We also fix \BERT and \SWIN (Section~\ref{sec:method:representation}) 
during adaptation for better efficiency.
After re-learning, 
for each user in the target task, we calculate the recommendation score of item $v_i$ as follows:
\begin{equation}
    \begin{aligned}
    \label{eqn:score:inductive}
    r_i &= \cos(\tilde{\mathbf{h}}^*, {\tilde{\mathbf{v}}_i^*}),
    \end{aligned}
\end{equation}
where $r_i$ is the recommendation score of item $v_i$.
We recommend the top-$k$ items of the highest recommendation scores.

\paragraph*{\method with Interaction-based Item Embedding (\methodT)}
During adaptation, we also develop an {\method} variant, denoted as \methodT, in which in addition to embeddings from item modalities (i.e., texts, images and prices), we learn another embedding $\tilde{\mathbf{e}}^*$ from user interactions for each item in the target task to generate 
target-task-specific item representations.
Thus, item embeddings in \methodT are as follows:
\begin{equation}
    \label{eqn:trans}
    \tilde{\mathbf{v}}_i^*  =
        \IRL(v_i^t, v_i^m, v_i^p | \boldsymbol{\Phi}_{\scriptsize{\IRL}}^*) +
        \tilde{\mathbf{e}}_i^*.
\end{equation}
In \methodT, we minimize $\mathcal{L}_r^*$ in terms of the interaction-based item embeddings, $\boldsymbol{\Phi}_{\scriptsize{\IRL}}^*$ 
and $\boldsymbol{\Theta}_{\scriptsize{\POS}}^*$
as follows:
\begin{equation}
    \label{eqn:loss:trans}
    \begin{aligned}
        \min \limits_{\boldsymbol{\Phi}_{\tiny{\IRL}}^*, E^*, \boldsymbol{\Theta}_{\scriptsize{\POS}}^*} \mathcal{L}_r^*, 
    \end{aligned}
\end{equation}
where $E^* = [\tilde{\mathbf{e}}_1^*;\tilde{\mathbf{e}}_2^*, \cdots]$ is a matrix containing the interaction-based embeddings for all the items in the target task. 
%

\section{Experimental Setup}
\label{sec:materials}

\subsection{Baseline Methods}
\label{sec:materials:baseline}

%
In our experiment, we compare \method with three state-of-the-art interaction-based baseline methods that generate item embeddings using only user interactions in the target task: 
1) \NPE~\cite{nguyen2018npe} uses neural networks to capture the non-linear relations among items;
2) \SASRec~\cite{kang2018self} estimates users' intent by leveraging the self-attention mechanisms;
3) \BERTRec~\cite{sun2019bert4rec} captures users' preferences using a bidirectional self-attention-based sequence encoder.

In addition, we compare \method with three state-of-the-art text-based baseline methods that generate item embeddings based on item texts, and user interactions:
1) \FDSA~\cite{zhang2019feature} uses two self-attention networks to capture the transitions among items, and the transitions among item texts;
2) \SRec~\cite{zhou2020s3} learns expressive item representations
by maximizing the mutual information between the embedding of items and item texts.
3) \UniSRec~\cite{hou2022towards} uses item texts to transfer knowledge across recommendation tasks.
Note that, \UniSRec is a pre-training-based method, and it has been demonstrated superior performance over all the other pre-training-based sequential recommendation methods such as \RECGUGU~\cite{li2022recguru} and \ZESREC~\cite{ding2021zero} 
by a significant margin.
Thus, we include \UniSRec for comparison instead of the methods that \UniSRec outperforms.
%
%

We also compare \method with the state-of-the-art multi-modality-based method \NOVA~\cite{liu2021noninvasive} and \SASRecP~\cite{rashed2022carca}, 
which leverage fusion operators to integrate item information from modalities.
Compared to \NOVA and \SASRecP, 
%
\method leverages pre-training and adaptation to transfer knowledge across tasks. 
%

For all the baseline methods except for \SASRecP and \NOVA, we use the implementation in RecBole~\footnote{\url{https://recbole.io/}}, a widely used library to benchmark recommendation methods.
For \SASRecP, In the original paper, only item texts and images are used.
To enable a fair comparison, we reimplement \SASRecP with PyTorch-Lighting~\footnote{\url{https://www.pytorchlightning.ai}} to incorporate item prices.
We also implement \NOVA since its implementation is not publicly available.
We use the same evaluation setting (Section~\ref{sec:materials:protocol}) 
and the same datasets for target tasks (Section~\ref{sec:materials:dataset}) 
in \method and all the baseline methods.
We also use the same text, image and price embeddings in \method and all the baseline methods if applicable.
%
%

We tune the hyper-parameters in \method and all the baseline methods using grid search. 
We use the same search range as reported in Hou~\etal~\cite{hou2022towards} to tune the hyper-parameters for the baseline methods.
We report the search ranges for each hyper-parameter, and other important implementation details in the Appendix.
%
%

\subsection{Datasets}
\label{sec:materials:dataset}

\begin{table}
  \caption{Dataset Statistics}
  \centering
  \label{tbl:dataset}
  \begin{threeparttable}
      \begin{tabular}{
	@{\hspace{3pt}}l@{\hspace{3pt}}
    @{\hspace{3pt}}l@{\hspace{3pt}}
	@{\hspace{3pt}}r@{\hspace{3pt}}          
	@{\hspace{3pt}}r@{\hspace{3pt}}
	@{\hspace{3pt}}r@{\hspace{3pt}}
	@{\hspace{3pt}}r@{\hspace{3pt}}
    @{\hspace{3pt}}r@{\hspace{3pt}}
	}
        \toprule
        task & dataset & \#users & \#items & \#intrns & {\#intrns/u} & \#u/i\\
        \midrule
        \multirow{4}{*}{Auxiliary}
        & {\Food}    &    115,349 &   39,670 &    1,027,413 &   8.9 & 25.9\\
        & \Movies &    281,700 &   59,203 &    3,226,731 & 11.5 & 54.5\\
        & \Home   &    731,913 & 185,552 &    6,451,926 &   8.8 & 34.8\\
        & \Cloth   &  1,164,752 & 372,593 & 10,714,172 &   9.2 & 28.8\\
        \midrule
        \multirow{5}{*}{Target}
        & \Scientific &      8,442 &     4,385 &        59,427 & 7.0 & 13.6\\
        & \Pantry    &    13,101 &     4,898 &       126,962 & 9.7 & 25.9\\
        & \Instruments & 24,962 & 9,964   &       208,926 & 8.4 & 21.0\\
        & \Arts        &    45,486 &    21,019 &      395,150  & 8.7 & 18.8\\
        & \Office     &    87,346 &    25,986 &      684,837  & 7.8 & 26.4\\
        \bottomrule
      \end{tabular}
      \begin{tablenotes}[normal,flushleft]
      \begin{footnotesize}
      \item 
      In this table, the column ``task" indicates if the datasets are used in auxiliary tasks or target tasks.
      The column ``\#users", ``\#items" and ``\#intrns" shows the number of 
      users, items and user-item interactions, respectively. 
      The column ``\#intrns/u" has the average number of interactions for each user. 
      The column ``\#u/i" has the average number of interactions for each item.  
      \par
      \end{footnotesize}
      \end{tablenotes}
  \end{threeparttable}
  \vspace{-10pt}
\end{table}

%
In our experiments, we consider four auxiliary tasks of
recommending items of the category food, movie, home and clothing for pre-training, 
and leverage the Amazon-Food (\Food), Amazon-Movies (\Movies), Amazon-Home (\Home) and Amazon-Clothing (\Cloth) datasets for the four tasks.
We select these four auxiliary tasks for pre-training because they contain a larger number of user interactions compared to the other tasks, 
and thus, could have more information for transfer.

In line with \UniSRec, we consider five target tasks of recommending items of the category scientific, pantry, instruments, arts and office. 
We use the Amazon-Scientific (\Scientific), Amazon-Pantry (\Pantry), Amazon-Instruments (\Instruments), Amazon-Arts (\Arts) and Amazon-Office (\Office) datasets for these target tasks.
%
All of the datasets are from Amazon reviews~\footnote{\url{https://jmcauley.ucsd.edu/data/amazon/}}, 
which include users' interactions and reviews on different categories of products.
%
%
To the best of our knowledge, publicly available datasets from sources other than Amazon reviews do not provide comprehensive item information (e.g., images and prices).
Thus, we only utilize datasets from Amazon reviews in our experiments.
Following \UniSRec, we only keep users and items with at least five interactions and concatenate 
title, sub-categories and brand of items as the item text.
In addition, for each user, we only consider her most recent 50 interactions in the experiments.
Table~\ref{tbl:dataset} presents the statistics of the processed datasets. 

\subsection{Experimental Protocol}
\label{sec:materials:protocol}

\subsubsection{Training, Validation and Testing Sets}
\label{sec:materials:protocol:training}

Following the literature~\cite{kang2018self,sun2019bert4rec,hou2022towards}, 
for each user, 
 we use her/his last interaction for testing, second last interaction for validation and all the other interactions for training.
We tune the hyper-parameters on the validation sets for all methods, 
and use the best-performing hyper-parameters in terms of Recall@$10$ as will be discussed below for testing.

\subsubsection{Evaluation Metrics}
\label{sec:materials:protocol:evaluation}

Following the literature~\cite{kang2018self,peng2020ham,ssr,hou2022towards}, 
we evaluate all the methods using Recall@$k$ and NDCG@$k$.
We refer the audience of interest to Peng~\etal~\cite{peng2020ham} for the detailed definitions of both Recall@$k$ and NDCG@$k$.
%
%
%
%
In the experiments, we report the average results over all the users for Recall@$k$ and NDCG@$k$.
%
%
We assess the statistical significance of performance differences at these evaluation metrics using the paired 
t-test.

\section{Experiments and Analysis}
\label{sec:exp}

\subsection{Overall Performance}
\label{sec:exp:overall}

\begin{table*}[!t]
\footnotesize
  \caption{Overall Performance}
  \centering
  \label{tbl:overall_performance}
  \begin{threeparttable}
      \begin{tabular}{
        @{\hspace{0pt}}l@{\hspace{3pt}}
	  @{\hspace{3pt}}l@{\hspace{4pt}}
	  @{\hspace{6pt}}r@{\hspace{5pt}}
	  @{\hspace{4pt}}r@{\hspace{3pt}}
	  @{\hspace{3pt}}r@{\hspace{4pt}}
	  @{\hspace{6pt}}r@{\hspace{4pt}}
        @{\hspace{6pt}}r@{\hspace{4pt}}
        @{\hspace{3pt}}r@{\hspace{3pt}}
        @{\hspace{2pt}}r@{\hspace{8pt}}
        @{\hspace{6pt}}r@{\hspace{2pt}}
        @{\hspace{1pt}}r@{\hspace{4pt}}
        @{\hspace{5pt}}r@{\hspace{3pt}}
        @{\hspace{5pt}}r@{\hspace{3pt}}
        @{\hspace{2pt}}r@{\hspace{0pt}}
      }
      \toprule
      Dataset & Metric & \NPE & \SASRec & \BERTRec & \FDSA & \SRec & \UniSRecI & \UniSRecT & \NOVA & \SASRecP & \method & \methodT & imprv (\%)\\
      \midrule
      \multirow{4}{*}{\Scientific}
      &  Recall@10 & 0.0772 & 0.1060 & 0.0500 & 0.0907 & 0.0500 & 0.1071 & 0.1163 & 0.0962 & \underline{0.1260} & 0.1334 & \textbf{0.1347}  & 6.9$^*$\\
      & Recall@50 & 0.1746 & 0.2022 & 0.1241 & 0.1777 & 0.1315 & 0.2172 & 0.2313 & 0.1761 & \underline{0.2362} & \textbf{0.2479} & 0.2460   & 5.0$^*$\\
      \cline{2-14}
      & NDCG@10   & 0.0375 & 0.0579 & 0.0254 & 0.0588 & 0.0255 & 0.0563 & 0.0613 & 0.0678 & \underline{0.0839} & 0.0864 & \textbf{0.0886} & 5.6$^*$\\
      & NDCG@50   & 0.0588 & 0.0787 & 0.0414 & 0.0777 & 0.0430 & 0.0802 & 0.0863 & 0.0851 & \underline{0.1079} & 0.1111 & \textbf{0.1130} & 4.7$^*$\\
      \midrule
      \multirow{4}{*}{\Pantry}
      &  Recall@10 & 0.0424 & 0.0537 & 0.0316 & 0.0437 & 0.0462 & 0.0647 & 0.0710 & 0.0536 & \underline{0.0811} & \textbf{0.0878} & 0.0864   & 8.3$^*$\\
      & Recall@50 & 0.1381 & 0.1385 & 0.1000 & 0.1314 & 0.1381  & 0.1719 & 0.1828 & 0.1284 & \underline{0.1886} & 0.1975 & \textbf{0.1983}   & 5.1$^*$\\
      \cline{2-14}
      & NDCG@10   & 0.0194 & 0.0235 & 0.0160 & 0.0233 & 0.0226 & 0.0312 & 0.0330 & 0.0288 & \underline{0.0446} & \textbf{0.0489} & 0.0485 & 9.6$^*$\\
      & NDCG@50   & 0.0400 & 0.0416 & 0.0306 & 0.0420 & 0.0423 & 0.0544 & 0.0571 & 0.0449 & \underline{0.0678} & 0.0727 & \textbf{0.0728} & 7.4$^*$\\
      \midrule
      \multirow{4}{*}{\Instruments}
      &  Recall@10 & 0.1008 & 0.1163 & 0.0821 & 0.1149 & 0.1119 & 0.1115 & 0.1260 & 0.1051 & \underline{0.1276} & 0.1340 & \textbf{0.1352}  & 6.0$^*$\\
      & Recall@50 & 0.1984 & 0.2125 & 0.1437 & 0.2116 & 0.2054  & 0.2107 & \underline{0.2418} & 0.1912 & 0.2301 & 0.2404 & \textbf{0.2439}  & 0.9\textcolor{white}{$^*$}\\
      \cline{2-14}
      & NDCG@10   & 0.0567 & 0.0673 & 0.0624 & 0.0756 & 0.0668 & 0.0642 & 0.0715 & 0.0766 & \underline{0.0924} & 0.0965 & \textbf{0.0969} & 4.9$^*$\\
      & NDCG@50   & 0.0778 & 0.0881 & 0.0756 & 0.0964 & 0.0870 & 0.0858 & 0.0967 & 0.0952 & \underline{0.1147} & 0.1196 & \textbf{0.1205} & 5.1$^*$\\
      \midrule
      \multirow{4}{*}{\Arts}
      &  Recall@10 & 0.0877 & 0.1122 & 0.0675 & 0.1034 & 0.1017 & 0.1016 & \underline{0.1278} & 0.1093 & 0.1195 & 0.1329 & \textbf{0.1361}  & 6.5$^*$\\
      & Recall@50 & 0.1790 & 0.2041 & 0.1313 & 0.1940 & 0.1929  & 0.1996 & \underline{0.2405} & 0.1941 & 0.2209 & 0.2370 & \textbf{0.2408}  & 0.1\textcolor{white}{$^*$}\\
      \cline{2-14}
      & NDCG@10   & 0.0467 & 0.0612 & 0.0433 & 0.0676 & 0.0567 & 0.0568 & 0.0700 & 0.0764 & \underline{0.0814}  & 0.0905 & \textbf{0.0938} & 15.2$^*$\\
      & NDCG@50   & 0.0666 & 0.0810 & 0.0571 & 0.0871 & 0.0765 & 0.0781 & 0.0945 & 0.0947 & \underline{0.1035}  & 0.1131 & \textbf{0.1166} & 12.7$^*$\\
      \midrule
      \multirow{4}{*}{\Office}
      &  Recall@10 & 0.0884 & 0.1186 & 0.0838 & 0.1134 & 0.1127 & 0.0996 & \underline{0.1269} & 0.1106 & 0.1178 & 0.1295 & \textbf{0.1326}   & 4.5$^*$\\
      & Recall@50 & 0.1527 & 0.1817 & 0.1237 & 0.1722 & 0.1751  & 0.1676 & \underline{0.2014} & 0.1683 & 0.1852 & 0.2021 & \textbf{0.2019}   & 0.2\textcolor{white}{$^*$}\\
      \cline{2-14}
      & NDCG@10   & 0.0502 & 0.0787 & 0.0635 & 0.0878 & 0.0775 & 0.0615 & 0.0800 & 0.0863 & \underline{0.0881} & 0.0969 & \textbf{0.1007} & 14.3$^*$\\
      & NDCG@50   & 0.0642 & 0.0925 & 0.0722 & 0.1006 & 0.0910 & 0.0763 & 0.0963 & 0.0988 & \underline{0.1028} & 0.1127 & \textbf{0.1158} & 12.6$^*$\\
      \bottomrule
      \end{tabular}
      \begin{tablenotes}[normal,flushleft]
      \begin{footnotesize}
      \item
      For each dataset, the best performance is in \textbf{bold}, 
      and the best performance among the baseline methods is \underline{underlined}.
      The column ``imprv" presents the percentage improvement of \method
      over the best-performing baseline methods (\underline{underlined}).
      The ${^*}$ indicates that the improvement is statistically significant at 90\% confidence level.
      \par
      \end{footnotesize}
      \end{tablenotes}
  \vspace{-10pt}
  \end{threeparttable}
\end{table*}


Table~\ref{tbl:overall_performance} presents the overall 
performance of \method, its variant \methodT and state-of-the-art baseline methods
at Recall@$k$ and NDCG@$k$ on the five datasets used for target tasks.
%
In the comparison, we also include a variant of \UniSRec, denoted as \UniSRecT.
Similarly to \methodT, \UniSRecT also learns interaction-based item embeddings to enable more task-specific item representations.

As shown in Table~\ref{tbl:overall_performance}, 
overall, \methodT is the best-performing method on the five datasets. 
In terms of Recall@$10$,
\methodT establishes the new state of the art on four out of five datasets (i.e., \Scientific, \Instruments, \Arts and \Office), and
achieves the second-best performance on \Pantry.
Similarly, at Recall@50, \methodT also outperforms all the other
methods on four out of five datasets except for \Scientific.
On average, over the five datasets, 
\methodT achieves a remarkable improvement of 6.1\% and 2.1\% at Recall@$10$ and Recall@$50$, respectively, compared to the best-performing baseline method at each dataset.
Similarly, in terms of NDCG@$10$ and NDCG@$50$, \method achieves statistically significant improvement over the best-performing baseline method across all the datasets.
Table~\ref{tbl:overall_performance} also presents that \methodI is the second-best performing method.
In terms of both Recall@$k$ and NDCG@$k$, \methodI achieves the best or second-best performance, 
and substantially outperforms the baseline methods on all the datasets. 
For example, in terms of Recall@$10$, \methodI shows a substantial average improvement of 5.1\% compared to best-performing baseline methods over the five datasets.
These results  
demonstrate the superior performance of \method over the state-of-the-art baseline methods.
%

\subsubsection{Comparison between \methodI and \methodT}
\label{sec:exp:overall:inductive}

We notice that as shown in Table~\ref{tbl:overall_performance}, without additional interaction-based item embeddings, the performance of 
\methodI is still highly competitive with that of \methodT.
For example, in terms of Recall@$10$, \method only slightly underperforms \methodT by 1.0\% averaged over the five datasets.
On \Pantry, \methodI even slightly outperforms \methodT at Recall@$10$.
These results suggest that with multi-modality information (i.e., texts, images and price) of items, 
in \methodI, we could generate comprehensive item embeddings.
As a result, 
adding interaction-based item embeddings may not provide significant additional information 
and thus, does not substantially benefit the recommendation performance.
To the best of our knowledge, \methodI is the first method demonstrating that interaction-based item embeddings are not necessary for the state-of-the-art recommendation performance, 
if comprehensive item embeddings as those in \method are employed. 
Note that, compared to \methodT, 
without interaction-based item embeddings,
\method could generate recommendations for new items which do not have interactions (i.e., zero-shot recommendation),
and thus, would be more practical in real applications.

\subsubsection{Comparison between \method and \SASRec}
\label{sec:exp:overall:ID}

%
Table~\ref{tbl:overall_performance} presents that 
\method achieves superior performance 
over the best-performing 
interaction-based method \SASRec on all the five datasets.
In terms of Recall@$10$ and NDCG@$10$, over the five datasets, \method significantly outperforms \SASRec 
with a remarkable average improvement of 26.4\% and 54.3\%, respectively.
The primary difference between \method and \SASRec is in two folds:
1) \method utilizes multi-modality item information to generate comprehensive item embeddings, while \SASRec learns item embeddings only using user interactions;
2) \method leverages the rich interaction data in auxiliary tasks to improve the recommendation performance in the target task via transfer learning, 
while \SASRec learns only from the sparse user interactions in the target task.
The substantial improvement of \method over \SASRec 
indicates the effectiveness of utilizing
multi-modality item information and transfer learning for recommendation. 
%

\subsubsection{Comparison between \method and \UniSRecT}
\label{sec:exp:overall:text}

%
As shown in Table~\ref{tbl:overall_performance}, 
\method also substantially outperforms the best-performing text-based method \UniSRecT at both Recall@$k$ and NDCG@$k$.
Specifically, compared to \UniSRecT, \method achieves a remarkable average improvement of 10.2\% and 34.9\% at Recall@$10$ and NDCG@$10$, respectively, 
over the five datasets.
As will be shown in Section~\ref{sec:exp:domain}, 
item texts could be substantially different across tasks.
Consequently, based on item texts, \UniSRecT could learn limited transferable knowledge on auxiliary tasks.
To address this limitation, \method also incorporates item images and item prices to more effectively learn transferable knowledge on auxiliary tasks. 
%

\subsubsection{Comparison between \method and \SASRecP}
\label{sec:exp:overall:carca}

Table~\ref{tbl:overall_performance} also shows that 
\method could consistently outperform the best-performing multi-modality-based method \SASRecP on all the datasets with substantial improvement (e.g., 5.9\% at Recall@$10$ on \Scientific).
In \method, we transfer knowledge from auxiliary tasks to the target task via pre-training and adaptation, 
while \SASRecP only learns from the information in the target task.
The improvement of \method over \SASRecP suggests the effectiveness of transfer learning in recommendation applications.

\subsection{Analysis on Negative Transfer Issue}
\label{sec:exp:transferibility}

\begin{table}[h]
\footnotesize
  \caption{Comparison between Adapted and Ground-built Models}
  \centering
  \vspace{-5pt}
  \label{tbl:ScratchVsPEFinetune}
  \begin{footnotesize}
  \begin{threeparttable}
      \begin{tabular}{
            @{\hspace{0pt}}l@{\hspace{2pt}}
            @{\hspace{2pt}}l@{\hspace{2pt}}
            @{\hspace{2pt}}c@{\hspace{2pt}}          
            @{\hspace{2pt}}c@{\hspace{2pt}}
            @{\hspace{2pt}}c@{\hspace{2pt}}
            @{\hspace{2pt}}c@{\hspace{2pt}}
            @{\hspace{2pt}}c@{\hspace{2pt}}
            @{\hspace{2pt}}c@{\hspace{2pt}}
            @{\hspace{2pt}}c@{\hspace{2pt}}
            @{\hspace{2pt}}c@{\hspace{0pt}}
      }
      \toprule
     \multirow{2}{*}{ Dataset} & \multirow{2}{*}{Metric} 
     && \multicolumn{3}{c}{\UniSRec} && \multicolumn{3}{c}{\method}\\
    \cline{4-6} \cline{8-10}
     &&& \SCRATCH & \FINETUNE &\DSA && \SCRATCH & \FINETUNE &\DSA\\
      \midrule
      \multirow{4}{*}{\Scientific}
      & Recall@10 && 0.1187 & 0.1071 & \textbf{0.1211} && 0.1299 & \textbf{0.1340} & 0.1334\\
      & Recall@50 && 0.2319 & 0.2172 & \textbf{0.2421} && 0.2406 & 0.2421 & \textbf{0.2479}\\
      \cline{2-10}
      & NDCG@10 && \textbf{0.0658} & 0.0563 & 0.0644 && 0.0846 & \textbf{0.0871} & 0.0864\\
      & NDCG@50 && 0.0905 & 0.0802 & \textbf{0.0908} && 0.1086 & 0.1106 & \textbf{0.1111}\\
      \midrule
      \multirow{4}{*}{\Pantry}
      & Recall@10 && 0.0698 & 0.0647 & \textbf{0.0711} && 0.0808 & 0.0859 & \textbf{0.0878}\\
      & Recall@50 && 0.1765 & 0.1719 & \textbf{0.1853} && 0.1933 & 0.1955 & \textbf{0.1975}\\
      \cline{2-10}
      & NDCG@10 && 0.0336 & 0.0312 & \textbf{0.0339} && 0.0444 & 0.0475 & \textbf{0.0489}\\
      & NDCG@50 && 0.0566 & 0.0544 & \textbf{0.0587} && 0.0687 & 0.0712 & \textbf{0.0727}\\
      \midrule
      \multirow{4}{*}{\Instruments}
      & Recall@10 && 0.1217 & 0.1115 & \textbf{0.1240} && 0.1305 & 0.1333 & \textbf{0.1340}\\
      & Recall@50 && 0.2232 & 0.2107 & \textbf{0.2311} && 0.2382 & 0.2356 & \textbf{0.2404}\\
      \cline{2-10}
      & NDCG@10 && \textbf{0.0729} & 0.0642 & 0.0723 && 0.0933 & 0.0960 & \textbf{0.0965}\\
      & NDCG@50 && 0.0949 & 0.0858 & \textbf{0.0955} && 0.1166 & 0.1183 & \textbf{0.1196}\\
      \midrule
      \multirow{4}{*}{\Arts}
      & Recall@10 && \textbf{0.1106} & 0.1016 & 0.1002 && 0.1284 & 0.1256 & \textbf{0.1329}\\
      & Recall@50 && \textbf{0.2122} & 0.1996 & 0.1990 && 0.2310 & 0.2282 & \textbf{0.2370}\\
      \cline{2-10}
      & NDCG@10 && \textbf{0.0641} & 0.0568 & 0.0557 && 0.0879 & 0.0855 & \textbf{0.0905}\\
      & NDCG@50 && \textbf{0.0861} & 0.0781 & 0.0772 && 0.1103 & 0.1078 & \textbf{0.1131}\\
      \midrule
      \multirow{4}{*}{\Office}
      & Recall@10 && \textbf{0.1055} & 0.0996 & 0.1037 && 0.1268 & 0.1248 & \textbf{0.1295}\\
      & Recall@50 && 0.1711 & 0.1676 & \textbf{0.1723} && 0.1982 & 0.1951 & \textbf{0.2021}\\
      \cline{2-10}
      & NDCG@10 && \textbf{0.0707} & 0.0615 & 0.0649 && 0.0949 & 0.0936 & \textbf{0.0969}\\
      & NDCG@50 && \textbf{0.0849} & 0.0763 & 0.0798 && 0.1104 & 0.1089 & \textbf{0.1127}\\
      \bottomrule
      \end{tabular}
      \begin{tablenotes}[normal,flushleft]
      \begin{footnotesize}
      \item
      In this table, \SCRATCH, \FINETUNE and \DSA represents ground-built models, fine-tuned models and models adapted with our relearning-based  strategy, respectively.
      For each dataset, the best performance in \UniSRec and \method are in \textbf{bold}.
      \par
      \end{footnotesize}
      \end{tablenotes}
  \end{threeparttable}
  \end{footnotesize}
  \vspace{-10pt}
\end{table}

We conduct an analysis
to evaluate if \UniSRec, the stat-of-the-art pre-training-based method, and \method suffer from the negative transfer issue.
Specifically, in this analysis, 
we compare the performance of ground-built models (i.e., models learned from scratch)
and adapted models (i.e., models adapted from the pre-trained model) of \UniSRec and \method in the target task.
We consider two adaptation strategies for learning adapted models: fine-tuning and re-learning in both \UniSRec and \method.
During fine-tuning, following that in \UniSRec~\cite{hou2022towards}, 
in both \UniSRec and \method, 
we fix the user intent model 
and only 
fine-tune the parameters for item representation learning.
For \UniSRec, we use the pre-trained model open-sourced by their authors.
%
We denote the ground-built model, the fine-tuned model 
and the re-learned model as \SCRATCH, \FINETUNE and \DSA, respectively.
We report the performance of \SCRATCH, \FINETUNE and \DSA from \UniSRec and \method on the five datasets in Table~\ref{tbl:ScratchVsPEFinetune}.
%

Table~\ref{tbl:ScratchVsPEFinetune} presents that, with fine-tuning, \UniSRec substantially suffers from the negative transfer issue.
Particularly, 
the fine-tuned model in \UniSRec, denoted as \UniSRec-\FINETUNE, underperforms the model built from scratch, denoted as \UniSRec-\SCRATCH, by a remarkable margin across all five datasets.
For example, at Recall@$10$, 
\UniSRec-\FINETUNE substantially underperforms 
\UniSRec-\SCRATCH by 8.5\% averaged over the five datasets.
%
By contrast, as shown in Table~\ref{tbl:ScratchVsPEFinetune} \method, when fine-tuned, is less susceptible to negative transfer. 
%
Specifically, 
\method-\FINETUNE outperforms 
the \method-\SCRATCH on \Scientific, \Pantry, and \Instruments with a considerable average improvement of 3.9\% at Recall@$10$. 
%
%

Table~\ref{tbl:ScratchVsPEFinetune} also presents that our re-learning-based adaptation strategy could substantially mitigate the negative transfer issue.
In particular, the relearned model in \UniSRec, denoted as \UniSRec-\DSA, outperforms \UniSRec-\SCRATCH on \Scientific, \Pantry and \Instruments with an average improvement of 1.9\% at Recall@$10$.
Similarly, \method-\DSA also consistently outperforms \method-\SCRATCH across all five datasets at all the evaluation metrics, and achieves a substantial average improvement of 3.9\% at Recall@$10$ over \method-\SCRATCH.
These results demonstrate that 
by incorporating multi-modality item information (Section~\ref{sec:method:representation})
and leveraging re-learning-based adaptation (Section~\ref{sec:method:finetuning}), \method could effectively address the negative transfer issue in existing methods.
\subsection{Analysis on Transferability}
\label{sec:exp:domain}
%

\begin{figure}[!h]
       \centering
       \footnotesize
        \begin{minipage}{\linewidth}
               \begin{subfigure}{0.32\linewidth}
                    \centering
                    \includegraphics[width=1.2\linewidth]{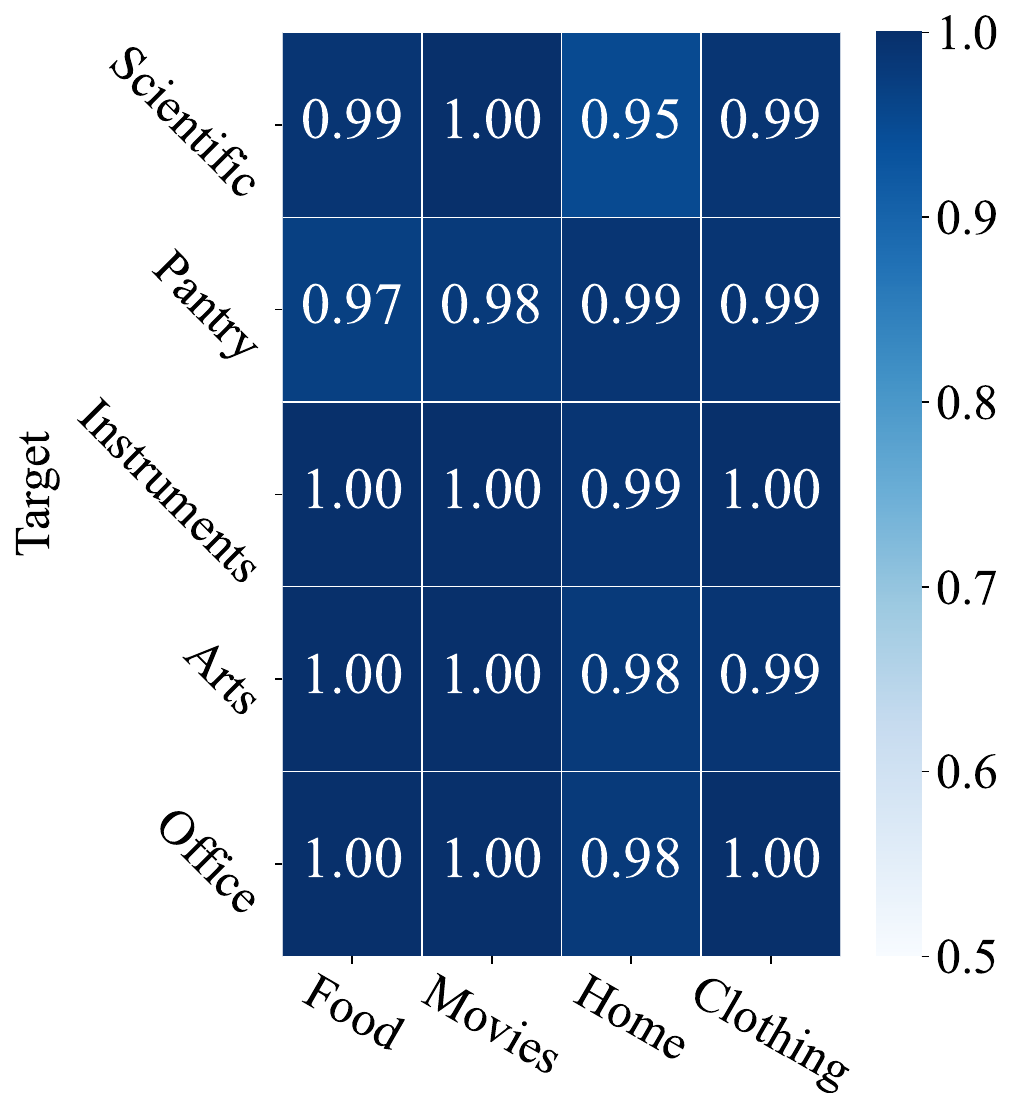}
                    \caption{Text}
                    \label{fig:text}
                \end{subfigure}
                \begin{subfigure}{0.32\linewidth}
                    \centering
                    \includegraphics[width=1.2\linewidth]{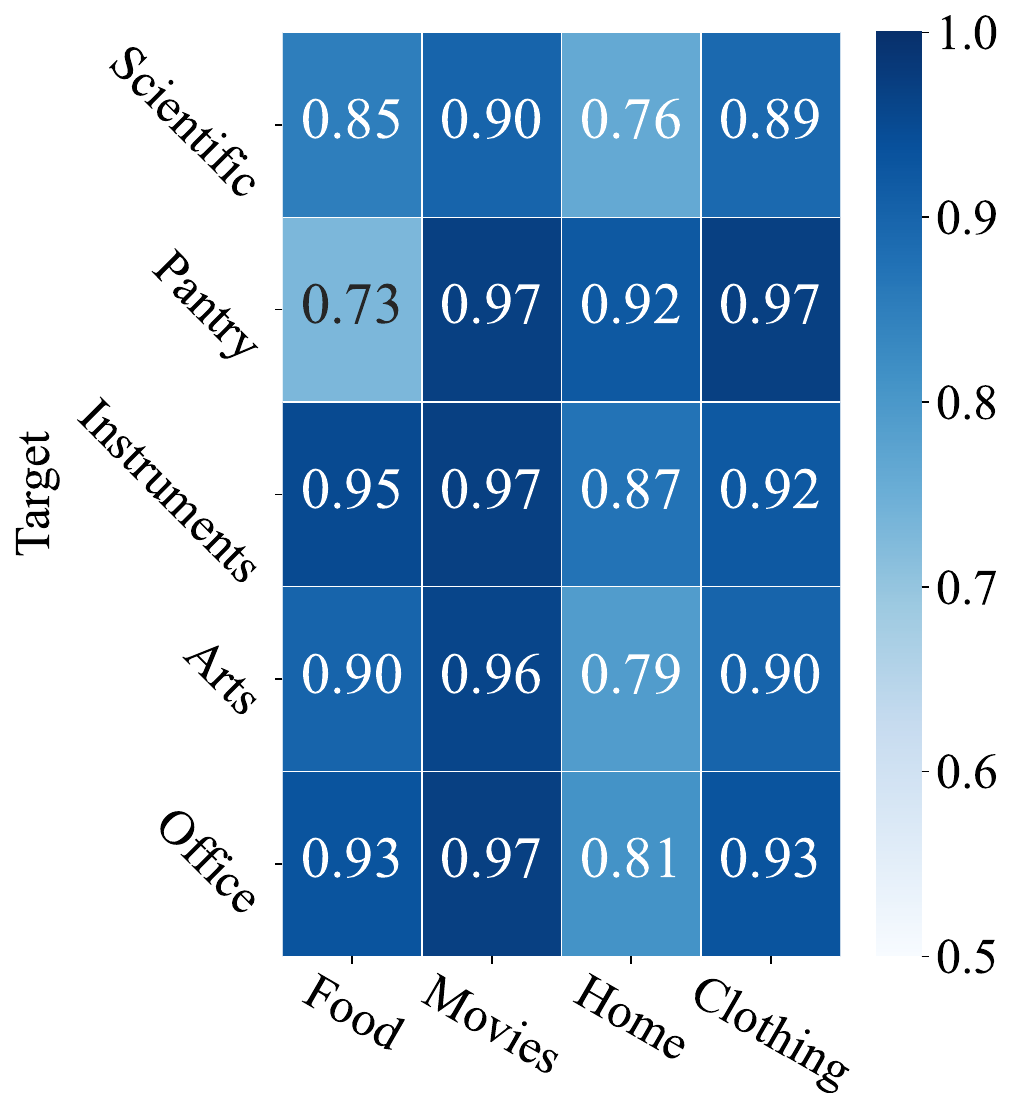}
                    \caption{Image}
                    \label{fig:image}
                \end{subfigure}
                \begin{subfigure}{0.32\linewidth}
                    \centering
                    \includegraphics[width=1.2\linewidth]{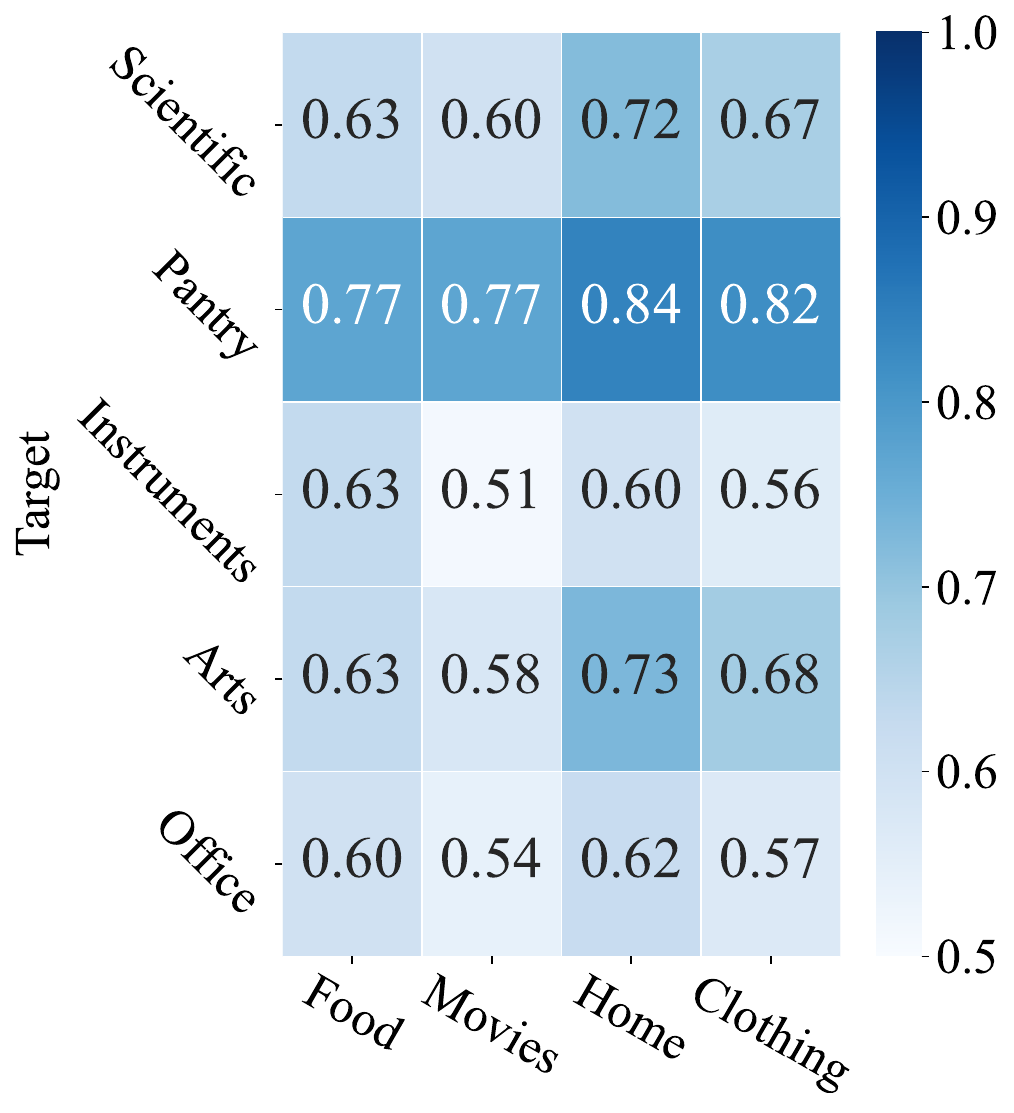}
                    \caption{Price}
                    \label{fig:price}
                \end{subfigure}
    \end{minipage}
\caption{Task Prediction Accuracy}
\label{fig:domain}
\vspace{-10pt}
\end{figure}

We conduct an analysis to investigate the cross-task transferability of item texts, images and prices. 
Intuitively, 
for each of the item modalities (e.g., text), 
we could assess its cross-task transferability using the task sensitivity of the modality embeddings (e.g., text embeddings).
The higher task sensitivity of the embeddings indicates lower cross-task transferability of the item modality.
%
%
Previous study~\cite{saito2018maximum} indicates that if the embeddings are task-sensitive, a classifier should be able to accurately separate the embeddings from different tasks.
Therefore, in this analysis, we measure the transferability of item texts, images and prices using the cross-task separability of their embeddings.

Particularly, in this analysis, for each target task,
we pair it with each of the four auxiliary tasks as in Table~\ref{tbl:dataset}.
For each of the pairs, given the embeddings of a certain item modality (e.g., texts, images and prices), 
we learn a linear binary classifier to separate the items in the target task from those in the auxiliary task.
We use the linear classifier since it has been widely used in studying the separability of embeddings~\cite{tian2020rethinking,saito2018maximum}.
To enable a balanced classification, for each item in the target task, 
we uniformly sample one item from the auxiliary task.
For items used for classification, 
we uniformly sample 70\%, 10\% and 20\% for training, validation and testing, respectively.
We use the accuracy on the validation set to select the best number of training epochs.

In this analysis, 
we use the transformed embedding $\tilde{\mathbf{z}}_i^t$, $\tilde{\mathbf{z}}_i^m$ and $\tilde{\mathbf{z}}_i^p$ (Equation~\ref{eqn:aggregate})
to represent the text, image and price of item $v_i$.
Given embeddings from a certain modality (e.g., text), 
we measure the cross-task separability of the embeddings using the accuracy of correctly predicting the task which has the corresponding items.
%
A higher accuracy indicates better separability, and thus lower cross-task transferability.

Figure~\ref{fig:text}, \ref{fig:image} and \ref{fig:price} 
shows the classification accuracy on each pair of tasks with text, image and price embeddings, respectively.
As shown in Figure~\ref{fig:domain}, 
text embeddings lead to the highest classification accuracy.
Specifically, with text embeddings, using simple linear classifiers, 
we could achieve accuracies of at least 0.95 on all the pairs 
indicating that the item texts in target tasks and auxiliary tasks could be significantly different.
Thus, item texts could suffer from limited cross-task transferability.
%
Figure~\ref{fig:domain} also shows that compared to text embeddings, the embeddings of item images and prices are less separable across tasks.
For example, in terms of item images, the classification accuracy on the pair (\Pantry, \Food) is 0.73 implying that items in \Pantry and \Food could have similar images.
Thus, knowledge transfer across the two tasks could be enabled via item images.
A similar trend could also be observed with item prices.
For item images and item prices, we also tried non-linear classifiers such as two-layer perceptrons but still get similar results (e.g., 0.78 on (\Pantry, \Food) with item image).
%
%
These results suggest that item images and prices exhibit stronger cross-task transferability compared to item texts.
Therefore, by incorporating them, 
\method could more effectively learn transferable knowledge on auxiliary tasks compared to existing pre-training-based methods.
%

%
%
%
%
%

\subsection{Ablation Study}
\label{sec:exp:ablation}

\begin{table}[!t]
\footnotesize
  \caption{Ablation Study}
  \centering
  \label{tbl:ablation}
  \begin{threeparttable}
      \begin{tabular}{
        @{\hspace{1pt}}l@{\hspace{1pt}}
        @{\hspace{1pt}}l@{\hspace{3pt}}
        @{\hspace{2pt}}c@{\hspace{3pt}}
        @{\hspace{2pt}}c@{\hspace{2pt}}
        @{\hspace{2pt}}c@{\hspace{2pt}}
        @{\hspace{2pt}}c@{\hspace{2pt}}
        @{\hspace{2pt}}c@{\hspace{2pt}}
        @{\hspace{2pt}}c@{\hspace{1pt}}
      }
      \toprule
      Dataset & Metric & \NoText & \NoImage & \NoPrice & \NoMoE & \NoFusion & \methodI\\
      \midrule
      
      \multirow{4}{*}{\Scientific}
      & Recall@10   & 0.1001 & 0.1253 & 0.1314 & 0.1288 & \textbf{0.1334} & \textbf{0.1334}\\
      & Recall@50   & 0.2000 & 0.2354 & \textbf{0.2510} & 0.2473 & 0.2441 & 0.2479\\
      \cline{2-8}
      & NDCG@10   & 0.0649 & 0.0813 & 0.0855 & 0.0840 & \textbf{0.0871} & 0.0864\\
      & NDCG@50   & 0.0866 & 0.1051 & \textbf{0.1115} & 0.1097 & 0.1112 & 0.1111\\
      \midrule
      \multirow{4}{*}{\Pantry}
      & Recall@10   & 0.0839 & 0.0776 & 0.0781 & 0.0845 & 0.0861 & \textbf{0.0878}\\
      & Recall@50   & 0.1856 & 0.1875 & 0.1851 & \textbf{0.1991} & 0.1965 & 0.1975\\
      \cline{2-8}
      & NDCG@10   & 0.0468 & 0.0425 & 0.0424 & 0.0473 & 0.0488 & \textbf{0.0489}\\
      & NDCG@50   & 0.0688 & 0.0665 & 0.0654 & 0.0719 & \textbf{0.0727} & \textbf{0.0727}\\
      \midrule
      \multirow{4}{*}{\Instruments}
      & Recall@10   & 0.1245 & 0.1222 & 0.1302 & 0.1294 & 0.1336 & \textbf{0.1340}\\
      & Recall@50   & 0.2127 & 0.2193 & 0.2366 & 0.2345 & 0.2379 & \textbf{0.2404}\\
      \cline{2-8}
      & NDCG@10   & 0.0898 & 0.0879 & 0.0931 & 0.0941 & 0.0958 & \textbf{0.0965}\\
      & NDCG@50   & 0.1088 & 0.1089 & 0.1162 & 0.1169 & 0.1184 & \textbf{0.1196}\\
      \midrule
      \multirow{4}{*}{\Arts}
      & Recall@10   & 0.1039 & 0.1159 & 0.1261 & 0.1215 & 0.1310 & \textbf{0.1329}\\
      & Recall@50   & 0.1785 & 0.2116 & 0.2299 & 0.2229 & 0.2359 & \textbf{0.2370}\\
      \cline{2-8}
      & NDCG@10   & 0.0736 & 0.0778 & 0.0860 & 0.0827 & 0.0894 & \textbf{0.0905}\\
      & NDCG@50   & 0.0899 & 0.0987 & 0.1085 & 0.1048 & 0.1123 & \textbf{0.1131}\\
      \midrule
      \multirow{4}{*}{\Office}
      & Recall@10   & 0.1123 & 0.1053 & 0.1056 & 0.1173 & 0.1262 & \textbf{0.1295}\\
      & Recall@50   & 0.1718 & 0.1705 & 0.1699 & 0.1851 & 0.1962 & \textbf{0.2021}\\
      \cline{2-8}
      & NDCG@10   & 0.0840 & 0.0772 & 0.0781 & 0.0877 & 0.0948 & \textbf{0.0969}\\
      & NDCG@50   & 0.0969 & 0.0914 & 0.0921 & 0.1024 & 0.1100 & \textbf{0.1127}\\
      \bottomrule
      \end{tabular}
      \begin{tablenotes}[normal,flushleft]
      \begin{footnotesize}
      \item
      In this table, ``{\NoText}", ``{\NoImage}", ``{\NoPrice}" , ``{\NoMoE}" and ``{\NoFusion}" represents 
      {\methodI} without item texts, item images, item prices, MoE and modality fusion respectively.
      The best performance in each dataset is in \textbf{bold}.
      \par
      \end{footnotesize}
      \end{tablenotes}
  \vspace{-15pt}
  \end{threeparttable}
\end{table}


We conduct an ablation study to investigate the contribution of 
item texts, item images, item prices, MoE-based embedding transformation 
and modality fusion for the recommendation performance in \method.
Specifically, to investigate the contribution of item texts, images and prices, we replace the text, image and price embeddings with zero vectors 
during the adaptation.
To investigate the contribution of MoE-based embedding transformation, 
we replace our MoE-based approach with simple linear projections for the transformation of text, image and price embeddings.
We also replace the embeddings modeling the synergy between item texts and images ($\mathbf{f}_i$ in Equation~\ref{eqn:fusion}) with zero vectors to investigate the contribution of modality fusion.
We denote the \methodI ablative variant without texts, images, prices, MoE-based embedding transformation and modality fusion 
as \NoText, \NoImage, \NoPrice, \NoMoE and \NoFusion, respectively, 
and present their performance in Table~\ref{tbl:ablation}.

As shown in Table~\ref{tbl:ablation}, with all the modules, 
\methodI outperforms all of its ablative variants on four out of the five datasets except for \Scientific.
On \Scientific, \methodI still achieves the best performance at Recall@$10$, 
and the second-best performance at Recall@$50$ and NDCG@$10$.
Particularly, at Recall@$10$, over the five datasets, 
\methodI achieves an average improvement of 17.8\%, 13.4\%, 9.0\%, 6.2\% and 1.3\% 
compared to \NoText, \NoImage, \NoPrice, \NoMoE and \NoFusion, respectively.
In terms of Recall@$50$, \methodI also substantially outperforms \NoText, \NoImage, \NoPrice, \NoMoE and \NoFusion 
with an average improvement of 18.8\%, 10.2\%, 5.8\%, 3.5\% and 1.3\%, respectively, over the five datasets.
A similar trend could also be observed at NDCG@$10$ and NDCG@$50$.
These results show that 
item texts, item images, item prices, MoE-based embedding transformation and modality fusion all significantly 
contribute to the recommendation performance of \method. 
%

\section{Conclusion}
\label{sec:conclusion}

In this paper, 
we show that state-of-the-art pre-training-based sequential recommendation methods substantially suffer from the negative transfer issue.
To address this issue, 
we introduce \method that integrates 
item texts, images and prices together 
to effectively learn 
transferable knowledge on auxiliary tasks.
In addition, \method employs a re-learning-based strategy to better capture the task-specific knowledge 
in the target task.
Our experimental results demonstrate that 
\method could remarkably outperform the eight state-of-the-art baseline methods, with an improvement of as much as 15.2\% in the five target tasks.
Our negative transfer analysis shows that \method could substantially mitigate the negative transfer issue, and the adapted \method model outperforms the ground-built one on the five target tasks.
Our transferability analysis shows that compared to item texts, item images and prices could carry more transferable knowledge.
Our ablation study indicates that item texts, images and prices all substantially contribute to the recommendation performance of \method.

\section*{Acknowledgment}
\label{sec:ack}

This project was made possible, in part, by support from the National Science Foundation
grant nos. IIS-2133650 (X.N.). Any opinions, findings and conclusions or recommendations
expressed in this paper are those of the authors and do not necessarily reflect the views of the funding agency. 

\bibliographystyle{IEEEtran}
\bibliography{main}

\vfill
\vspace*{-3\baselineskip}

%
\begin{IEEEbiographynophoto}{Bo~Peng}
is a Ph.D. student at the Computer
Science and Engineering Department, The Ohio State University.
His research interests include machine learning, data mining and their applications in
recommender systems and graph mining.
\end{IEEEbiographynophoto}
\vspace*{-2.5\baselineskip}
\begin{IEEEbiographynophoto}{Srinivasan~Parthasarathy}
received his Ph.D. degree from the Department of Computer Science,
University of Rochester, Rochester, in 1999. He is currently 
a Professor at the Computer Science and Engineering Department, The Ohio State University.
His research is on high-performance data analytics, graph analytics and network science, and machine learning and database systems.

\end{IEEEbiographynophoto}
\vspace*{-2.5\baselineskip}
\begin{IEEEbiographynophoto}{Xia~Ning}
received her Ph.D. degree from the Department of Computer Science and Engineering,
University of Minnesota, Twin Cities, in 2012. She is currently 
a Professor at the Biomedical Informatics Department, and the Computer
Science and Engineering Department, The Ohio State University.
Her research is on data mining, machine learning and artificial intelligence with applications 
in recommender systems, drug discovery and medical informatics. 
\end{IEEEbiographynophoto}
\vfill

\clearpage

\appendices

\setcounter{table}{0}
\renewcommand{\thetable}{A\arabic{table}}

\section{Reproducibility}
\label{app:implementation}

\begin{table*}
\footnotesize
  \caption{\mbox{Best-performing Hyper-parameters in \method, \methodT and Baseline Methods}}
  \centering
  \label{tbl:para}
  \begin{threeparttable}
      \begin{tabular}{
        @{\hspace{8pt}}l@{\hspace{8pt}}
        @{\hspace{4pt}}r@{\hspace{4pt}}
        @{\hspace{4pt}}r@{\hspace{4pt}}
        @{\hspace{8pt}}c@{\hspace{8pt}}
        @{\hspace{4pt}}r@{\hspace{4pt}}
        @{\hspace{4pt}}r@{\hspace{4pt}}
        @{\hspace{8pt}}c@{\hspace{8pt}}
        @{\hspace{4pt}}r@{\hspace{4pt}}
        @{\hspace{4pt}}r@{\hspace{4pt}}
        @{\hspace{8pt}}c@{\hspace{8pt}}
        @{\hspace{4pt}}r@{\hspace{4pt}}
        @{\hspace{4pt}}r@{\hspace{4pt}}
        @{\hspace{8pt}}c@{\hspace{8pt}}
        @{\hspace{4pt}}r@{\hspace{4pt}}
        @{\hspace{4pt}}r@{\hspace{4pt}}
        @{\hspace{8pt}}c@{\hspace{8pt}}
        @{\hspace{4pt}}c@{\hspace{4pt}}
        @{\hspace{8pt}}c@{\hspace{8pt}}
        @{\hspace{4pt}}c@{\hspace{4pt}}
        @{\hspace{8pt}}c@{\hspace{8pt}}
        @{\hspace{4pt}}r@{\hspace{4pt}}
        @{\hspace{8pt}}c@{\hspace{8pt}}
        @{\hspace{4pt}}c@{\hspace{4pt}}
        }
        \toprule
        \multirow{2}{*}{Dataset} 
        & \multicolumn{2}{c}{\NPE}
        && \multicolumn{2}{c}{\SASRec}  
        && \multicolumn{2}{c}{\BERTRec} 
        && \multicolumn{2}{c}{\FDSA} 
        && \multicolumn{2}{c}{\SRec} 
        && \multicolumn{1}{c}{\NOVA} 
        && \multicolumn{1}{c}{\SASRecP} 
        && \multicolumn{1}{c}{\method} 
        && \multicolumn{1}{c}{\methodT}\\
        \cmidrule(lr){2-3}
        \cmidrule(lr){5-6} 
        \cmidrule(lr){8-9} 
        \cmidrule(lr){11-12} 
        \cmidrule(lr){14-15} 
        \cmidrule(lr){17-17}
        \cmidrule(lr){19-19} 
        \cmidrule(lr){21-21}
        \cmidrule(lr){23-23}
        & $d$ & $lr$
	  && $d$ & $lr$
        && $d$ & $lr$
        && $d$ & $lr$
        && $d$ & $lr$
        && $d$
        && $d$
        && $\beta$
        && $\beta$\\
        \midrule
        \Scientific & 300 & 3e-3
        && 128 & 1e-3  
        && 128 & 1e-2 
        && 64 & 3e-3 
        && 300 & 3e-3
        && 256
        && 256 
        && 0.1 
        && 0.1\\ 
	  \Pantry	& 128 & 3e-3
        && 300 & 3e-4  
        && 300 & 3e-3 
        && 64 & 1e-3 
        && 300 & 1e-3
        && 256
        && 256
        && 0.3 
        && 0.5\\
	  \Instruments & 128 & 1e-3
        && 300 & 3e-4  
        && 300 & 3e-4 
        && 128 & 3e-3    
        && 300 & 1e-3
        && 128
        && 256 
        && 0.7 
        && 0.1\\
	  \Arts & 128 & 1e-3
        && 300 & 3e-4  
        && 300 & 3e-4 
        && 128 & 3e-3 
        && 300 & 3e-3
        && 256
        && 256 
        && 0.7 
        && 0.1\\
	  \Office & 128 & 1e-3
        && 300 & 3e-4  
        && 300 & 3e-4 
        && 128 & 1e-3 
        && 300 & 3e-3
        && 128
        && 256 
        && 0.5 
        && 0.1\\
        \bottomrule
      \end{tabular}
      \begin{tablenotes}[normal,flushleft]
      \begin{scriptsize}
      \item
      This table presents the best-performing hyper-parameters of \method, \methodT and all the baseline methods except for \UniSRec on the five datasets.
      \par
      \end{scriptsize}
      \end{tablenotes}
      \vspace{-10pt}
  \end{threeparttable}
\end{table*}

We implement \method in Python 3.9.13 with PyTorch 1.10.2~\footnote{\url{https://pytorch.org/}} and PyTorch Lightning 1.7.7~\footnote{\url{https://www.pytorchlightning.ai/}}. 
Following \UniSRec and \SASRec, 
we set the number of self-attention layers and the number of attention heads in each layer as 2, and the dimension for item embeddings as 256 for \method and all the baseline methods if applicable.
We also use the cross entropy loss for \method and all the baseline methods to enable a fair comparison.
Following \UniSRec, in \method, we set the scaling hyper-parameters $\tau$ (Equation~\ref{eqn:rec}) as 0.07; 
set the number of projection heads in embedding transformation $n_h$ as 8; 
and use Adam optimizer with learning rate 1e-3 on all the datasets.
In \method, we also set the dimension of price embeddings $d_p$ as 64; 
use 50,000 as the frequency coefficient $\omega$ (Equation~\ref{eqn:price});
and normalize the item prices to be in the range of $[0, 100]$.
Considering the memory limits in GPUs,
we use batch optimization during pre-training and adaptation, and set the batch size as 2,048 following \UniSRec.
For \method and all the baseline methods, we initialize all the learnable parameters using the default initialization methods in PyTorch.
For \NOVA and \SASRecP, we search the dimension of item embeddings $d$ from $\{64, 128, 256\}$.
In \method, we search the coefficient of modality interaction embeddings $\beta$ (Equation~\ref{eqn:item}) from $\{0.1, 0.3, 0.5, 0.7\}$.
For \UniSRec, we use the hyper-parameters reported by the authors in all the datasets.
%
For \NPE, \SASRec, \BERTRec, \FDSA and \SRec, following \UniSRec, we search $d$ and the learning rate $lr$ 
from $\{64, 128, 300\}$ and $\{3e\text{-}4, 1e\text{-}3, 3e\text{-}3, 1e\text{-}2\}$, respectively.
%
%
Table~\ref{tbl:para} presents the best-performing hyper-parameters in \method, \methodT and all the baseline methods except for \UniSRec on the five datasets.
We refer the audience of interest to the \UniSRec's GitHub~\footnote{\url{https://github.com/RUCAIBox/UniSRec}} for the hyper-parameters used in \UniSRec.

\end{document}